\documentclass[utd,final,nonblindrev]{utd_V1}

\OneAndAHalfSpacedXI
\usepackage{tabularx}
\usepackage{multirow}
\usepackage{bm} 
\usepackage{graphicx}
\usepackage{subfigure, epsfig}
\usepackage{natbib}
\usepackage{verbatim}
 \bibpunct[, ]{(}{)}{,}{a}{}{,}%
 \def\bibfont{\small}%
\TheoremsNumberedThrough
\ECRepeatTheorems
\newtheorem{observation}{Observation}
\EquationsNumberedThrough
\begin{document}
\RUNTITLE{How to Promote Autonomous Driving}
\TITLE{How to Promote Autonomous Driving with Evolving Technology: Business  Strategy and Pricing Decision}

\ARTICLEAUTHORS{%
\AUTHOR{Mingliang Li}
\AFF{Antai College of Economics and Management, Shanghai Jiao Tong University, Shanghai, China\\
\EMAIL{limingliang@sjtu.edu.cn}} %, \URL{}}
\AUTHOR{Yanrong Li}
\AFF{Department of Industrial Systems Engineering \& Management,
National University of Singapore, Singapore\\
\EMAIL{liyr@nus.edu.sg}}
\AUTHOR{Lai Wei}
\AFF{Antai College of Economics and Management, Shanghai Jiao Tong University Shanghai, China \\
\EMAIL{laiwei@sjtu.edu.cn}}
\AUTHOR{Wei Jiang}
\AFF{Antai College of Economics and Management, Shanghai Jiao Tong University, Shanghai, China\\
\EMAIL{jiangwei@sjtu.edu.cn}}
\AUTHOR{Zuo-Jun Max Shen}
\AFF{Faculty of Engineering and Faculty of Business and Economics, The University of Hong Kong, Hong Kong Special Administrative Region, China\\
College of Engineering, University of California Berkeley, Berkeley, CA, USA\\
\EMAIL{vp-research@hku.hk}}
% Enter all authors
} % end of the block
\ABSTRACT{Recently, autonomous driving system (ADS) has been widely adopted due to its potential to enhance travel convenience and alleviate traffic congestion, thereby improving the driving experience for consumers and creating lucrative opportunities for manufacturers.
With the advancement of data sensing and control technologies, the reliability of ADS and the purchase intentions of consumers are continually evolving, presenting challenges for manufacturers in promotion and pricing decisions.
To address this issue, we develop a two-stage game-theoretical model to characterize the decision-making processes of manufacturers and consumers before and after a technology upgrade.
Considering the unique structural characteristics of ADS, which consists of driving software and its supporting hardware (SSH), we propose different business strategies for SSH (bundle or unbundle with the vehicle) and driving software (perpetual licensing or subscription) from the manufacturer's perspective. 
We derive the optimal software and SSH strategies and propose several managerial insights related to their interactions. 
First, SSH strategies influence the optimal software strategies by changing the consumers' entry barriers to the ADS market.
Specifically, for manufacturers with mature ADS technology, the bundle strategy provides consumers with a lower entry barrier by integrating SSH, making the flexible subscription model a dominant strategy; while perpetual licensing outperforms under the unbundle strategy.
Second, the software strategies influence the optimal SSH strategy by altering consumers' exit barriers.
Perpetual licensing imposes higher exit barriers; when combined with a bundle strategy that lowers entry barriers, it becomes a more advantageous choice for manufacturers with mature ADS technology.
In contrast, the subscription strategy allows consumers to easily exit the market, making the bundle strategy advantageous only when a substantial proportion of consumers are compatible with ADS.
Finally, we find that, when a manufacturer has rapidly improved technologies, perpetual licensing is always a better choice.}


\maketitle

\section{Introduction}\label{section:introduction}
With the continuous advancement of autonomous driving system (ADS) technology, its integration and application are increasingly recognized by both automotive manufacturers and consumers, as it has the potential to reduce accident rates, alleviate driver workloads, and improve transportation efficiency.
Currently, ADS, as an emerging add-on feature in traditional vehicles, has attracted increasing investments.
According to \cite{fortune2023autonomous}, the global autonomous vehicle market was valued at \$1,921 billion in 2023, and is projected to grow at an annual rate of 32.3\% and reach \$13,632 billion by 2030.
Numerous automotive manufacturers are progressively launching their proprietary autonomous driving technologies and offering them to consumers.
For example, Tesla has recently introduced the V12 version of their Full Self-Driving (FSD) service;\footnote{https://www.notateslaapp.com/news/2307/tesla-fsd-v1256-introduces-major-improvements-a-look-at-all-the-changes-photos} Mercedes-Benz has introduced the world's first Level-3 autonomous driving system.\footnote{https://media.mbusa.com/releases/automated-driving-revolution-mercedes-benz-announces-us-availability-of-drive-pilot-the-worlds-first-certified-sae-level-3-system-for-the-us-market}
These advancements present manufacturers with unprecedented opportunities while simultaneously challenging them to innovate their associated business models.
Therefore, developing the corresponding sales strategies that accommodate the continuously improving ADS is crucial for manufacturers.

Although extensive research has made significant contributions to add-on sales strategies from the manufacturer's point of view \citep{allon2011would,lin2017add,shugan2017product,cui2018unbundling,geng2018add}, these strategies cannot be applied directly to ADS for two reasons: the dynamic nature of consumers' willingness to adopt and the complexity of strategies arising from the unique structure of ADS, which are detailed below.

Consumers' willingness to adopt ADS depends on their perceptions of the technology's \textit{compatibility} and \textit{reliability}, both of which exhibit dynamics.
Compatibility is a prerequisite for consumers before making a purchase decision. 
As an emerging technology that has yet to achieve widespread adoption, autonomous driving may encounter resistance from a segment of conservative consumers who take a cautious attitude toward new technologies and perceive it as incompatible with their vehicle usage habits.
However, these perceptions may initially be inaccurate and exhibit dynamics as consumers gain more information about the technology and functionality of ADS \citep{jalili2020pricing, bai2023implementing}.
For consumers who perceived ADS as compatible, reliability is another critical factor influencing their purchasing decision.
With the development of autonomous driving technology, the reliability improvements of ADS offer potential purchasing opportunities for consumers who are reluctant to adopt the technology due to its initial reliability falling below their expectations. As a result, consumers' willingness to adopt ADS is also subject to temporal dynamics.
However, to the best of our knowledge, the dynamic perceptions of consumers regarding compatibility and reliability have not been considered in the design of add-on business strategies in the existing literature.

\begin{figure}[!ht]
\centering
\includegraphics[width=0.8\textwidth]{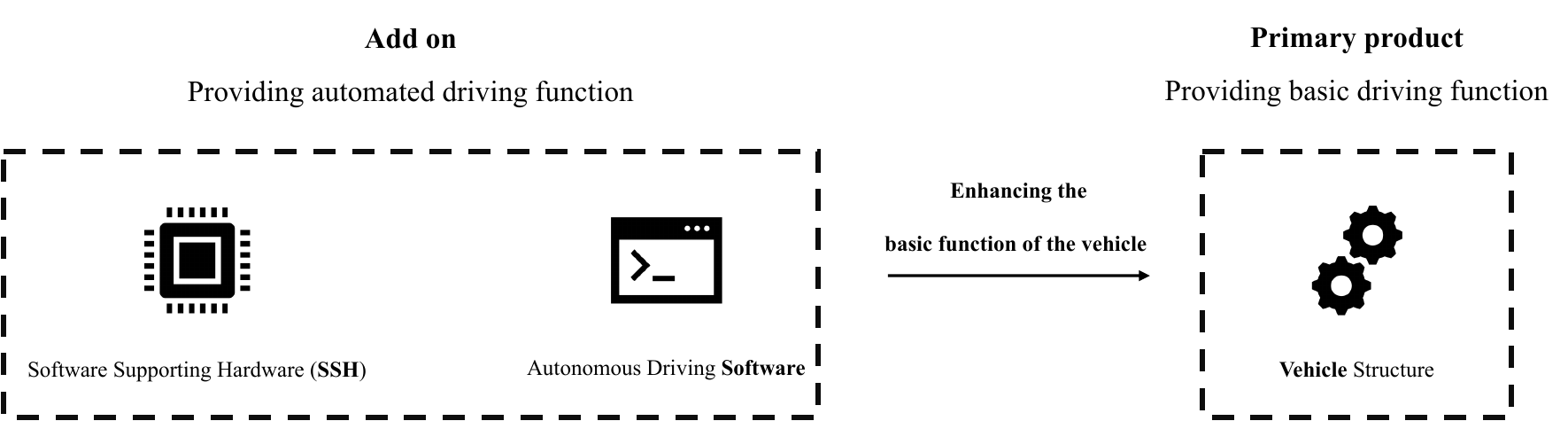} 
\caption{Three components of autonomous driving vehicles}
\label{background}
\end{figure}

Another key reason for the differences in ADS strategies is its structure.
The functionality of ADS depends not only on software, but also on a set of software support hardware (SSH) such as radars and chips (as shown in Figure \ref{background}). 
The inherent distinction between tangible SSH and digital software results in significant differences in their business strategies.
Due to this unique structure, manufacturers must consider combinations of hardware and software strategies for effective promotion, and existing add-on strategies cannot be directly applied in ADS.

\begin{itemize}
    \item For SSH strategies, manufacturers can bundle SSH with the vehicle (adopted by Tesla and NIO)\footnote{Tesla is the largest EV manufacturer in North America, while NIO, Xpeng, BYD, and Li Auto are prominent EV manufacturers in China} or sell them unbundle (adopted by Xpeng and Li Auto).
    Under the bundle strategy, manufacturers install SSH in every vehicle, allowing consumers to activate ADS at their discretion.
    However, under the unbundle strategy, consumers must determine whether to install SSH when purchasing a vehicle based on their perceived compatibility and reliability of ADS. Notably, SSH cannot be retrofitted after the vehicles are sold. Consumers who initially choose not to install SSH will permanently lose the opportunity to use ADS.
    \item For software strategies, there are also two options for manufacturers: perpetual licensing (adopted by Li Auto, Xpeng, and BYD) and subscription (adopted by NIO) strategies. Under the perpetual licensing strategy, consumers can always access the most recent software version once they purchase the license. In contrast, the subscription strategy allows consumers to make periodic payments only when they choose to use ADS.
\end{itemize}
In summary, although SSH and software strategies are widely adopted in practice, clear industry standards and criteria for strategic selection have not yet been established.
Therefore, this study aims to fill this gap and provide optimal strategies for different automotive manufacturers.

To address the aforementioned two unique characteristics in the design of ADS business strategies, we establish a game-analytical framework between the manufacturer and strategic consumers to derive optimal strategies for the manufacturer to maximize its profit. 
To fully characterize the dynamics in consumers' perceived compatibility and reliability of ADS, we propose a two-stage decision process to describe their behaviors before and after ADS upgrades.
Then, by combining strategies for SSH (i.e., \underline{u}nbundle or \underline{b}undle) and software (i.e., \underline{p}erpetual licensing or \underline{s}ubscription), we propose four potential strategies for manufacturers ($UP$, $US$, $BP$, and $BS$) and obtain the corresponding equilibrium profits and optimal prices.
We theoretically examine the dominant conditions for each strategy and analyze the interaction effects between the strategies of SSH and software.
Our findings provide criteria for strategy selection and offer the corresponding managerial insights as follows.

First, SSH strategies influence the optimal software strategy by altering the entry barriers of consumers to the ADS market. 
The unbundle strategy requires consumers to make separate decisions about purchasing SSH, thus limiting market entry for consumers with lower interest in ADS.
In contrast, the bundle strategy integrates SSH and provides all consumers with the opportunity to enter the market with a lower entry barrier.
This distinction significantly increases the heterogeneity of consumer behavior in the market under the bundle strategy by attracting consumers with diverse characteristics.
The rise in consumer heterogeneity, in turn, enhances the advantages of subscription for manufacturers with mature ADS technology, enabling them to employ more flexible subscription options to capture extra consumer surplus. However, the unbundle strategy allows only a subset of consumers who choose to install SSH to enter the market, resulting in a more homogeneous pattern of consumer behavior and diminishing the advantage of subscription. Consequently, manufacturers who choose an unbundle strategy prefer perpetual licensing.

Second, we find that software strategies influence optimal SSH strategies by changing exit barriers of consumers. 
Perpetual licensing provides a higher exit barrier for consumers and makes it easier for manufacturers to retain overall profits by expanding software sales.
As long as the ADS delivers sufficient utility, manufacturers will prefer the bundle strategy with a lower entry barrier combined with the perpetual licensing strategy to attract larger potential consumer groups. 
In contrast, under a subscription strategy with lower exit barriers, consumers can cancel their subscriptions if they find ADS incompatible.
Consequently, the additional consumers gained through bundling may be insufficient to compensate for the losses incurred when subscriptions are canceled. 
Additionally, manufacturers should bear the cost of pre-installing SSH under the bundle strategy. 
Consequently, manufacturers offering subscriptions typically prefer the unbundle strategy, which allows them to charge separately for SSH.
Only when the proportion of consumers compatible with ADS is sufficiently large, and the number of consumers canceling their subscriptions is relatively low, manufacturers consider adopting a bundle strategy to further expand the market.

Finally, we find that the magnitude of the technological improvement of ADS has a particularly significant impact on software strategies. 
If a manufacturer that implements a significant upgrade to ADS technology adopts a subscription strategy, then consumers will subscribe only after the ADS upgrade, resulting in homogeneous behaviors that, in turn, reduce the advantages of the subscription strategy.
Consequently, manufacturers whose ADS technology improves significantly are more inclined to adopt a perpetual licensing strategy.

The rest of this paper is organized as follows.
Section~\ref{section:literature_review} reviews literature related to our work.
Section~\ref{section:model_setting} introduces the problem statement and the model formulation.
In Section~\ref{section:model_analysis}, we discuss the equilibrium decisions and key properties of the four proposed strategies ($UP$, $US$, $BP$, and $BS$).
Section~\ref{section:discussion} gives the main result of this study, where we propose criteria for manufacturers' strategy selection through profit comparison and discuss the interaction between the SSH strategy and software strategies.
Section~\ref{section:extension} provides an extension based on different upgrade magnitudes of ADS reliability to obtain more general strategy selection rules.
Finally, Section~\ref{section:conclusion} presents the conclusions, managerial implications, and some future research directions.
All proofs are provided in Appendix \ref{appendix:proof_lemma} and \ref{appendix:proof_proposition}.
\section{Literature review}\label{section:literature_review}
Our work is primarily related to two streams of research: add-ons and the selling or leasing of quality-improving durable products.
We review the literature on these two aspects and position our work by combining and analyzing the existing literature as follows, then highlighting our specific contributions.

Add-on products, or ancillary products, are typically products or services that enhance the functionality of a primary product.
Several studies focus on whether firms should bundle their primary products with add-ons or sell them separately.
\cite{allon2011would} found that unbundling add-ons from primary products reduces the firm's costs and enhances social efficiency.
\cite{cui2018unbundling} focus on the impact of heterogeneous consumers on the bundle strategies of primary products and add-ons.
Firms decide whether to bundle add-ons and engage in price discrimination based on the fraction of high-type consumers who value the add-on service. 
\cite{lin2017add} supplements the research of \cite{allon2011would} by extending it to a competitive market environment of high-end and low-end firms.
The study revealed that high-end firms are more inclined to sell add-ons separately. 
In contrast, low-end firms tend to offer ancillary products for free when the cost-to-value ratio of the ancillary is sufficiently small.
\cite{shugan2017product} analyzed the bundling strategies of different types of product lines. 
Their findings indicate that bundling ancillary products with high-end offerings is more advantageous for product lines with low core differentiation.
Conversely, bundling the add-on with low-end offerings is more beneficial for product lines with high core differentiation.

Other studies have considered the impact of supply channels on the primary product and add-on strategies.
\cite{geng2018add} investigated how the distribution contracts of downstream companies influence the bundling strategies of downstream companies for add-on products.
They found that upstream firms prefer bundling add-on products with core products under wholesale contracts but opt to retail add-on products separately under agency contracts.
\cite{gao2022free} examined whether firms should offer free add-on products when distributing their main product through various channels.
They observed that when the main product is distributed through a single channel, and there is minimal differentiation in product quality compared to competitors, the firms are inclined to offer free add-ons.
Simultaneously, when services are provided through multiple channels, the cost of add-on products significantly impacts firms' strategic decisions across different channels.
\cite{zhang2023manufacturer} studied the scenarios where two primary products of different quality are provided through two different channels: the rental channel and the sales channel.
Their research highlighted that varying degrees of quality differentiation within the product line and the value of ancillary products play pivotal roles in shaping the firm's decision-making process.
Unlike \cite{zhang2023manufacturer} considering the supply method of the primary product, \cite{wang2019pricing} analyzed the optimal profits and consumer welfare of firms offering add-on services through different methods: subscription and pay-per-use models.
In their models, the quality of the add-on product is typically assumed to remain constant.
However, in the context examined in this study, the quality of the autonomous driving system improves continuously over time.
Unlike previous research, we investigate the optimal decision-making of the company when the quality of the add-on product evolves over time. Furthermore, we explore the impact of the product’s initial quality and the extent of its upgrades on the choice of strategies.

The earliest systematic explanation of the issue of selling or leasing durable products comes from a hypothesis proposed by \cite{coase1972durability}.
When a company sells durable products across different periods, the products introduced in later periods may compete with those sold earlier.
Based on this, \cite{coase1972durability} suggested that it is more profitable for a monopolistic enterprise to lease durable products rather than sell them.
\cite{stokey1981rational} tested Coase's hypothesis in both continuous and discrete time settings, finding that the hypothesis holds in continuous time.
However, in discrete time, the conclusions are influenced by the duration of each period.
\cite{bucovetsky1986concurrent} expanded Coase's hypothesis in a competitive environment, and they came to a different conclusion that selling is more profitable than leasing in this scenario.
\cite{chien2008sale} found that when network effects are present, the profits from selling durable products may be higher than those from leasing.

In the literature on selling or leasing durable goods, the studies most closely related to our research are those by \cite{choudhary1998economic,choudhary2007comparison,zhang2010perpetual,jia2018selling} and \cite{wang2023better}, which compare the selling and leasing strategies of quality-improving software.
\cite{choudhary1998economic} was the first to discuss the issues of software selling and leasing in the context of software upgrades.
Their findings support Coase's hypothesis that software leasing helps monopolistic companies increase market share and creates a win-win situation where firms increase revenue, and consumers gain greater welfare.
Moreover, \cite{choudhary2007comparison} found that companies are willing to provide higher quality software through the subscription model, further enhancing the user experience.
The study by \cite{zhang2010perpetual} integrated considerations of the uncertainty of quality upgrades and network effects.
They found that companies tend to prefer the subscription model in situations with high uncertainty about the quality of future software versions. Simultaneously, companies may lean towards the sales model when software demonstrates significant network effects.
\cite{jia2018selling} conducted a study comparing the profit outcomes of firms choosing between subscription and selling models under various price discrimination strategies.
They suggest that, under the selling model, firms should engage in price discrimination based on consumers' historical purchase behavior, whereas inter-temporal price discrimination is more advantageous for firms in the subscription model.
\cite{wang2023better} extended the research of \cite{jia2018selling} to a duopolistic competitive business environment and found that the decisions of competing firms would change with variations in the magnitude of upgrades.
Their models focus solely on the sales strategy of the software product, without considering the hardware that supports the software. 
In contrast, our research focuses on both hardware and software sales strategies and explores how the hardware strategy influences the choice of software strategy (between perpetual licensing and subscription).

In summary, by analyzing limitations of existing research on add-ons and the selling or leasing of quality-improving durable products, we fill the research gap by demonstrating the optimal strategy when the add-on is ADS, which consists of two components of hardware and software and its reliability improves over time.
At the same time, the interactive effects and their underlying mechanisms between the choice of add-on strategies for hardware components and software are emphasized. 
\section{Problem description and model formulation}\label{section:model_setting}
Due to sequential innovations in technology, the reliability of ADS will improve steadily. These improvements are verified by data collected from the autonomous vehicle tester (AVT) program.\footnote{https://www.dmv.ca.gov/portal/vehicle-industry-services/autonomous-vehicles/disengagement-reports/} 
When quantifying the reliability of autonomous systems using disengagement data, which refers to situations where the system encounters scenarios it cannot handle and requires manual intervention to ensure safety, we have the following analytical results in Table~\ref{Data_verification}. As shown in Table~\ref{Data_verification}, the disengagement rate decreases by approximately 29\% from 2022 to 2023.
Taking into account this improvement in ADS reliability, we develop a two-stage model to characterize consumer behaviors before and after technological upgrades \citep{dhebar1994durable,krishnan2011integrated,lim2015toward}.
Specifically, in Stage 1, the manufacturer releases its initial version of ADS with reliability $q$. In Stage 2, an upgraded version with higher reliability $\gamma q$ is provided, where $\gamma$ represents the magnitude of the upgrade to quantify the improvement in reliability. We set $\gamma$ to 1.3 in Sections~\ref{section:model_analysis} and \ref{section:discussion} based on Table~\ref{Data_verification}, and relax this assumption in Section~\ref{section:extension}.
Since pre-announcing the release of new technology is a strategy that can enhance a manufacturer's competitiveness \citep{lobel2016optimizing}, we assume that $q$ and $\gamma$ are disclosed to consumers at the beginning of Stage 1.

\begin{table}[!ht] 
    \caption{Data verification on reliability improvement}
    \fontsize{10pt}{10pt}\selectfont 
    \centering 
    {\def\arraystretch{1.5}
  \begin{tabular*}{\textwidth}{@{\extracolsep{\fill}}cccc}
\hline
\hline
  Test year & \# of disengagement events  & \# of kilometers & Disengagement rate/ million kilometer\\
\hline
  2022 & 8,216 & 5.1 million & 1,611 \\
  2023 & 6.562 & 5.7 million & 1,142 \\
\hline
\hline
    \end{tabular*}
    }
    \label{Data_verification}
\end{table}

Based on the aforementioned model setting, we elaborate on the detailed strategies of manufacturers and consumers in Subsections~\ref{subsec:3_manufacture_intro} and \ref{subsec:3_consumer_intro}, respectively, along with the sequence of their decision-making process in Subsectionw~\ref{subsec:3_sequence}.
\subsection{The manufacturer}\label{subsec:3_manufacture_intro}
As illustrated in Figure~\ref{background}, the functionality of ADS depends both on SSH and on software.
Manufacturers have two potential strategies for SSH: selling the vehicle structure and SSH as a bundle or selling them separately (unbundle).
Then, the cost of the vehicle (denoted as $c_v$) and the cost of SSH (denoted as $c_h$) should be considered in our model.
The decision variable is the bundle price $p_b$ under the bundle strategy, whereas, under the unbundle strategy, the decision variables are the vehicle price $p_v$ and SSH $p_h$.
In practice, the manufacturer sells SSH in conjunction with vehicles only in Stage 1, while software services are available to consumers in both stages.
The reason is that software can be delivered to consumers at any time through over-the-air (OTA) technology, while SSH is a complete hardware system that cannot be retrofitted.
Hence, under the unbundle strategy, consumers must decide in Stage 1 whether to install SSH at the time of vehicle purchase, which is a prerequisite for utilizing ADS.
Furthermore, since reliability improvements are delivered through OTA software updates, any consumer who has SSH retains the potential to access the latest version of ADS.

For software strategies, the manufacturer has two potential options: perpetual licensing and subscription \citep{zhang2010perpetual,jia2018selling,wang2023better}. 
Under the perpetual licensing strategy, consumers can use the latest version of ADS at every stage once they purchase the license at price $p_s$.   
We adopt this setting because most automotive manufacturers typically do not charge extra fees for reliability upgrades in practice, and the theories proposed by \citet{brecko2023new} also support this setting.
Under the subscription strategy, consumers must pay the subscription fee $r_s$ at each stage to access the latest version of ADS service. 

To briefly express different strategies, we denote the \underline{b}undle strategy for SSH and vehicle while offering software services through \underline{p}erpetual licensing as $BP$; denote the strategy of selling vehicles and SSH as a \underline{b}undle while offering software subscriptions as $BS$. Similarly, we use $UP$ and $US$ to denote \underline{p}erpetual licensing and \underline{s}ubscriptions under \underline{u}nbundle strategy.

\subsection{The consumer}\label{subsec:3_consumer_intro}
We assume that all consumers are strategic in estimating the costs and utilities of using ADS \citep{chen2015recent,bai2023implementing}, and make a two-stage decision based on the manufacturer's pricing strategy and the compatibility of the ADS both before and after reliability upgrade.
Suppose that consumers' perceived utility for using a vehicle only in each stage is $v$, and the manufacturing cost is less than the total utility in two stages ($2v > c_v$).
This condition must be satisfied because the vehicle's price must be set between $c_v$ and $2v$ to simultaneously ensure consumers' willingness to pay and the manufacturer's profitability.
Consumers' purchasing decisions depend on their evaluations of ADS, which are influenced by two key factors: prerequisite condition\textemdash the compatibility between themselves and ADS; subsequent condition\textemdash reliability of ADS.

Compatibility with ADS is a prerequisite for consumers before making a purchase decision, which is related to their driving behaviors and personal preferences for novel autonomous driving technologies. At the beginning of Stage 1, based on their prior assessment of compatibility, consumers can be categorized into two types: progressive consumers, who are highly receptive to new technologies and believe ADS is suitable for their vehicle usage scenarios; and conservative consumers, who take a cautious attitude toward new technologies and do not consider ADS suitable for them. Without loss of generality, we assume that each consumer type is equally represented and normalized to 1.
However, due to the limited information available at the beginning of Stage 1, consumers' initial assessments may not be precise \citep{jiang2017p2p, jalili2020pricing, bai2023implementing}.
Subsequently, more information, such as first-hand (e.g., personal usage experience) or second-hand (e.g., feedback from other users or media coverage), is collected during the usage in Stage 1 and will help consumers reassess their compatibility with ADS. 
As a result, more accurate posterior assessments are achieved irrespective of whether they have purchased ADS in Stage 1.  Suppose that both progressive and conservative consumers will perceive the ADS as compatible (incompatible) with the probability $\alpha$ ($1-\alpha$) in their posterior assessments. Based on the posterior assessment, we define the consumers who finally find the system compatible as system-compatible consumers; otherwise, they are denoted as system-incompatible consumers.
Figure 2 provides a brief schematic illustration of different types of consumers.
\begin{figure}[!ht]
    \centering
    \label{figure:consumer_type}
    \includegraphics[width=0.8\textwidth]{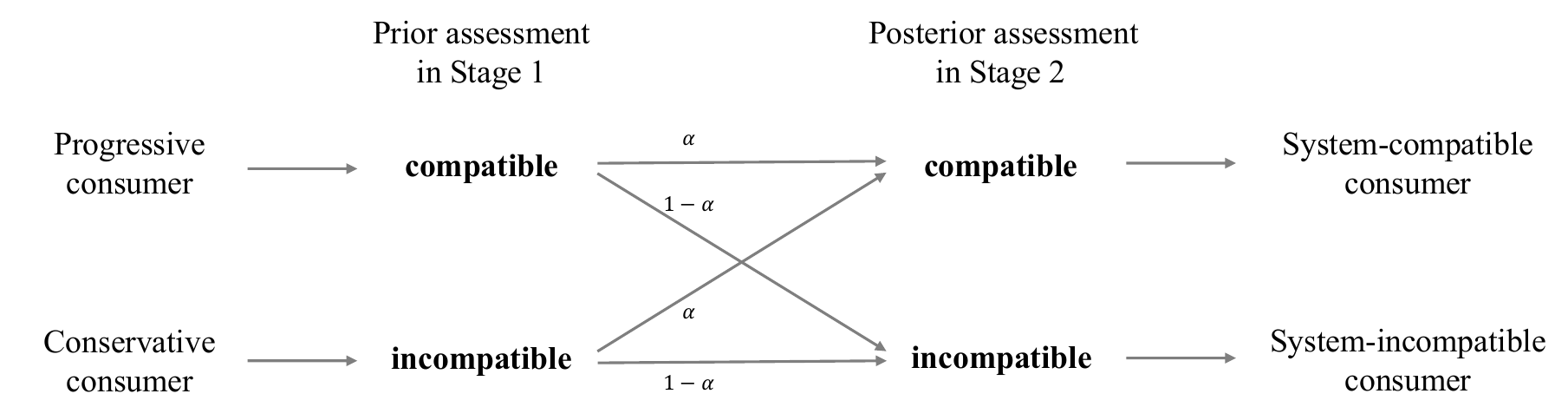}
    \caption{Illustration of different types of consumers}
\end{figure}

It is important to note that only consumers who perceive the ADS as compatible (i.e., progressive consumers in Stage 1 and system-compatible consumers in Stage 2) have a probability of paying for it, and their final decision depends on their perceived reliability of ADS. Consumers generally have heterogeneity and exhibit varying perceived reliability of ADS, which is characterized by a parameter $\theta$ following a uniform distribution $\theta\sim U(0,1)$.
If a consumer chooses to use ADS, it will have utility $v \cdot \theta q$ in Stage 1 and $v \cdot \theta \gamma q$ in Stage 2; otherwise, the utility is $v$ in both stages. 
We assume that $q > 1$ holds as we focus primarily on the enhancement of the overall utility provided by ADS. 
In summary, Table~\ref{Notation_table} presents all the aforementioned notations.

\begin{table}[!ht] 
    \caption{Notations}
    \fontsize{10pt}{10pt}\selectfont 
    \centering 
    {\def\arraystretch{1.5}
  \begin{tabular*}{\textwidth}{@{\extracolsep{\fill}}ccl}
\hline
\hline
    &Symbol&Definition\\
\hline
    \multirow{7}{*}{\begin{tabular}{@{}c@{}} Exogenous \\ variables\end{tabular}}
     & $\theta$&Consumers' perceived reliability of ADS, $\theta\sim U[0,1]$\\
    & $q$ &  Initial reliability of ADS in Stage 1, $q>1$\\
    & $\gamma$ & Upgrade level of ADS reliability, $\gamma=1.3$\\
     & $\alpha$& Proportion of system-compatible consumer, $0< \alpha<1$\\
    & $v$& The usage utility of the vehicle structure in each stage\\
    & $c_v$&Cost of the vehicle structure, $0<c_v<2v$\\
    & $c_h$&Cost of SSH, $c_h>0$\\
                \hline
    \multirow{5}{*}{\begin{tabular}{@{}c@{}} Decision \\ variables\end{tabular}}
    & $p_s$&Selling price of autonomous driving software\\
    & $r_s$&Subscription price of autonomous driving software in each stage\\
    & $p_h$ & Selling price of SSH\\
    & $p_v$ & Selling price of the vehicle structure\\
    & $p_b$ & Selling price of the bundle of vehicle structure and SSH\\
\hline
\hline
    \end{tabular*}
    }
    \label{Notation_table}
\end{table}

\subsection{Sequence of event}\label{subsec:3_sequence}

Suppose that the manufacturer first determines the strategies for SSH and software, as well as the corresponding prices. The consumers then make purchase decisions in two stages. In Stage 1, consumers decide whether to purchase the vehicle and SSH, as well as software, according to their prior assessment of the compatibility and perceived reliability of ADS. Subsequently, in Stage 2, consumers will adjust their purchasing or subscribing behaviors based on the posterior assessments.
Figure 3 illustrates this decision sequence.

\begin{figure}[!ht]
    \centering
    \label{figure:event}
    \includegraphics[width=0.8\textwidth]{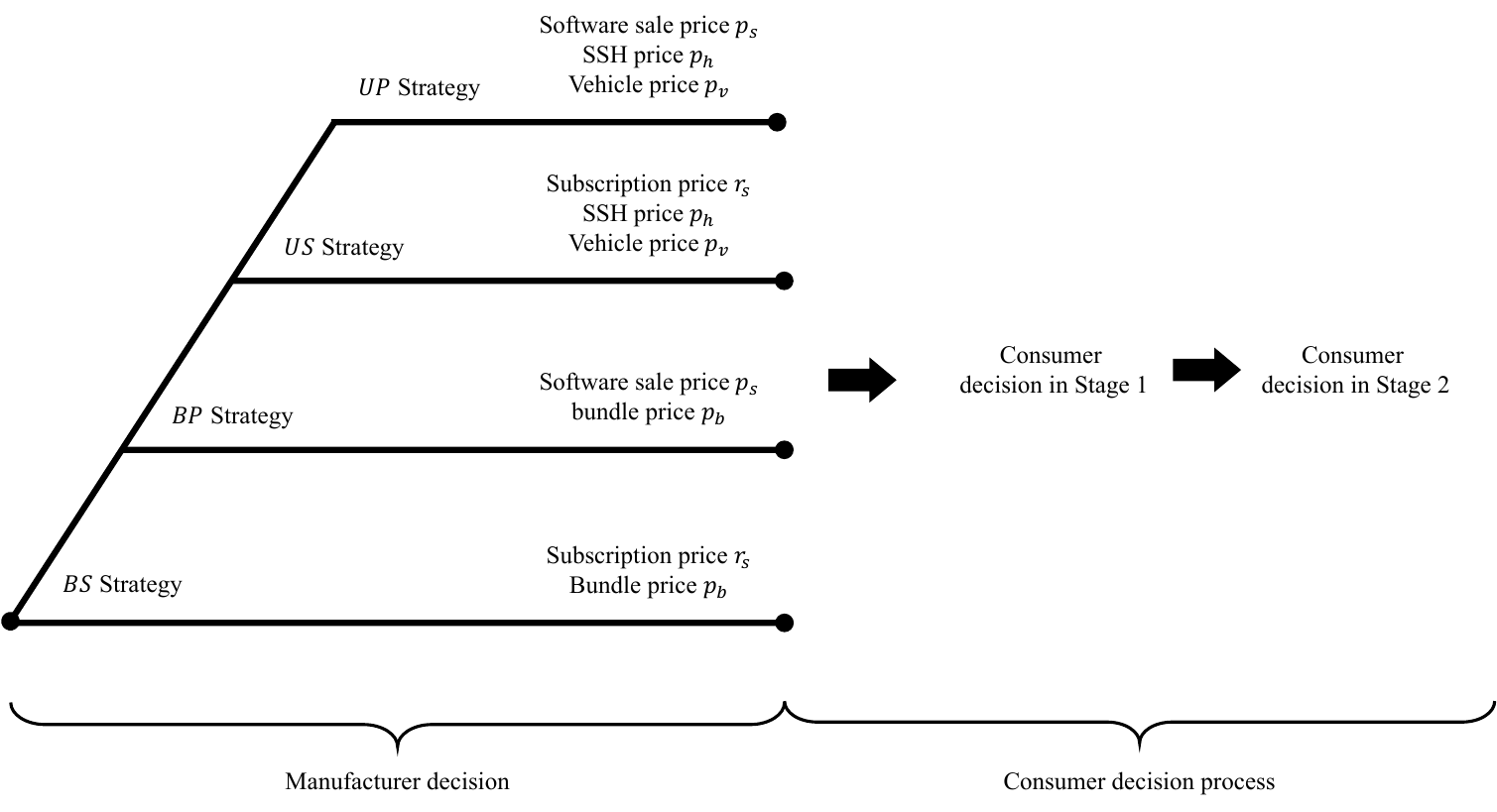}
    \caption{The sequence of events}
\end{figure}
\section{Equilibrium analysis}\label{section:model_analysis}
In this section, we examine the optimal approaches for both the manufacturer and consumers within a game-theoretic framework, and then independently derive the equilibrium solutions for the manufacturer under the $UP$, $US$, $BP$, and $BS$ strategies.
As introduced in Section \ref{subsec:3_consumer_intro}, the basic utility of using the vehicle over two stages is $2v$. If the vehicle's price is less than or equal to $2v$, all consumers will purchase the vehicle regardless of whether they intend to use ADS.
This scenario is considered in subsections \ref{subsection:4_1} to \ref{subsection:4_4}.
When the vehicle's price exceeds $2v$, consumers who do not intend to use ADS are unlikely to purchase it.
This scenario is discussed in subsection \ref{subsection:4_5}.
All proofs of the Lemmas~\ref{lemma:lemma1} to \ref{lemma:lemma6} are provided in the Appendix~\ref{appendix:proof_lemma}.

\subsection{$UP$ strategy}\label{subsection:4_1}
\begin{figure}[htbp]
\centering
\includegraphics[width=0.7\textwidth]{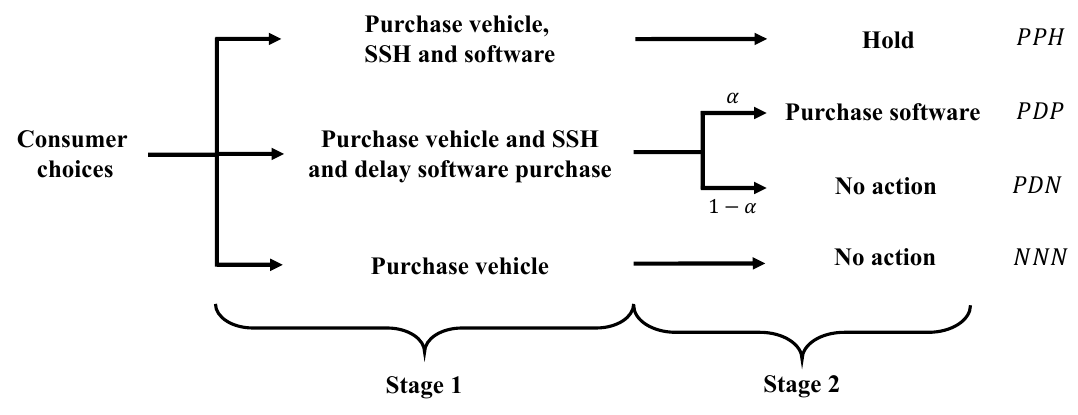}
\caption{Consumer choice under the $UP$ strategy}
\label{sequence_U_P}
\end{figure}

Under the $UP$ strategy, the manufacturer sells the vehicle at price $p_v$ and SSH at price $p_h$, along with a perpetual software license priced at $p_s$.
Progressive consumers who perceive ADS as compatible with their vehicle usage scenarios have three potential forward-looking behaviors for SSH and software in Stage 1: (1) use ADS in two stages: \underline{p}urchasing the SSH, \underline{p}urchasing software in Stage 1, and \underline{h}olding them in Stage 2, is denoted as $PPH$; (2) use ADS only after it is upgraded: \underline{p}urchasing SSH but \underline{d}elaying software purchase in Stage 1, and \underline{p}urchasing software in Stage 2, is denoted as $PDP$; (3) not to use ADS in both stages: \underline{n}ot purchasing SSH, \underline{n}ot purchasing software in Stage 1, and \underline{n}o action in Stage 2, is denoted as $NNN$.
In contrast, all conservative consumers believe that ADS is ineffective and choose the $NNN$ behavior in Stage 1. 
In Stage 2, consumers who opted for $PPH$ and $NNN$ cannot alter their decisions, as perpetual licensing for the software is non-refundable, and SSH cannot be retrofitted onto the vehicle.
However, consumers who choose to opt $PDP$ in Stage 1 can revise their behavior based on the posterior assessment of the compatibility with ADS.
Among these consumers, only a proportion $\alpha$ will continue with the behavior $PDP$ and purchase the software in Stage 2. The remaining $1-\alpha$ proportion of consumers will find that ADS is incompatible and choose \underline{n}ot to proceed with software purchase, we denote this behavior as $PDN$.
Based on different behavior choices, we divide consumers into different groups, which is shown in Figure~\ref{sequence_U_P}.

Next, we analyze the utility of consumers adopting different behaviors. If consumers choose $PPH$, their expected utility is  $v \theta q-p_s-p_v-p_h$ in Stage 1 and $v \theta\gamma q$ in Stage 2. The total utility function in two stages is $U^{PPH}=v \theta (q +\gamma q)-p_s-p_v-p_h$.
Similarly, the expected total utility functions based on the behaviors $PDP$ and $NNN$ are $U^{PDP}=v+v \theta \gamma q -p_s-p_v-p_h$, and $U^{NNN}=2v-p_v$, respectively.

One primary factor influencing consumer behavior is the heterogeneity in the perceived reliability of ADS ($\theta$).
We find that all utility functions ($U^{PPH}$, $U^{PDP}$, $U^{NNN}$) are linear and non-decreasing in $\theta$, and the condition
$$\frac{\partial U^{PPH}}{\partial \theta}>\frac{\partial U^{PDP}}{\partial \theta}>\frac{\partial U^{NNN}}{\partial \theta},$$
holds.
Therefore, as $\theta$ increases, the consumer's preference for the three behaviors changes in the following order at the beginning of Stage 1: $PPH\succ PDP\succ NNN$. 
We denote $\theta_{12}^{UP}$ as the marginal perceived reliability of consumers who are indifferent between the behaviors $PPH$ and $PDP$. When $\theta > \theta_{12}^{UP}$, consumers prefer $PPH$ to $PDP$. Similarly, the indifferent point between the behaviors $PDP$ and $NNN$ ($PPH$ and $NNN$) is denoted as $\theta_{23}^{UP}$ ($\theta_{13}^{UP}$).  

Based on the utilities, we can determine the number of consumers choosing different behaviors, which in turn allows us to derive the demand for ADS.
If consumers choose $PPH$, which indicates that they achieve the highest utility under $PPH$ behavior, the condition $\text{max}\left\{\theta_{12}^{UP}, \theta_{13}^{UP}\right\} < \theta < 1$ is satisfied. 
Then, we see that the number of consumers choosing $PPH$ is $D_{PPH}^{UP}=\text{max}\left\{0,1-\text{max}\left\{\theta_{12}^{UP},\theta_{23}^{UP}\right\}\right\}$.

Similarly, the number of consumers choosing $PDP$ and $PDN$ is $D_{PDP}^{UP}=\alpha \cdot \text{max}\left\{0,\theta_{12}^{UP}-\theta_{23}^{UP}\right\}$ and $D_{PDN}^{UP}=(1-\alpha) \cdot \text{max}\left\{0,\theta_{12}^{UP}-\theta_{23}^{UP}\right\}$, respectively, where $\alpha$ and $1-\alpha$ represent the proportions of system-compatible and system-incompatible consumers.
More detailed derivations on the value of indifferent points are shown in Table~\ref{indifferent_point_UB}.

Then, we obtain the total software demand as $D_{PPH}^{UP}+D_{PDP}^{UP}$ and the SSH demand is $D_{PPH}^{UP}+D_{PDP}^{UP}+D_{PDN}^{UP}$.
Since all consumers purchase the vehicle, the vehicle demand is equal to 2.
By maximizing the following profit function of the manufacturer, 
\begin{equation}
    \pi^{UP}\left(p_s,p_h,p_v\right)= \left(D_{PPH}^{UP}+D_{PDP}^{UP}\right)p_s+\left(D_{PPH}^{UP}+D_{PDP}^{UP}+D_{PDN}^{UP}\right)(p_h-c_h)+2(p_v-c_v),
\end{equation}
we have Lemma~\ref{lemma:lemma1}, and the closed-form solutions are presented in Table~\ref{proof:lemma1}.

\begin{lemma} \label{lemma:lemma1}
Under $UP$ strategy, the manufacturer will always package the sale of software with SSH.
\end{lemma}

This lemma suggests that when the manufacturer does not bundle the vehicle with SSH, the optimal choice is to package the SSH with the software for sale.
It is important to note that this package sale does not affect the behavior of consumers selecting $PPH$ and $NNN$, as they would purchase both components or do not purchase either in the first place, respectively.
The difference introduced by package sales is primarily reflected among consumers who choose $PDP$.
If software and SSH are sold separately, a proportion $1-\alpha$ of $PDP$ consumers will transition to $PDN$ consumers and abandon the software in Stage 2, resulting in a loss of profits for the manufacturer.
When software and SSH are packaged together, $PDP$ consumers are required to purchase the package in Stage 1 and cannot abandon the software in Stage 2 even if they find it incompatible.
Hence, this packaging approach is a superior choice for the manufacturer by preventing the emergence of $PDN$ consumers and mitigating profit losses.

\subsection{$US$ strategy}\label{subsection:4_2}
Under the $US$ strategy, the manufacturer sells the vehicle (by price $p_v$) and SSH (by price  $p_h$) separately and provides the software as a subscription service (by price $r_s$).
Progressive consumers have three forward-looking behaviors for SSH and software in Stage 1: (1) use ADS in two stages: \underline{p}urchasing the SSH, \underline{s}ubscribing software in Stages 1, and \underline{s}ubscribing in Stages 2, is denoted as $PSS$; (2) use ADS only after it upgraded: \underline{p}urchasing the SSH but \underline{d}elaying subscription the in Stage 1, and \underline{s}ubscribing in Stage 2, is denoted as $PDS$; (3) not to use ADS in both stages: \underline{n}ot purchasing SSH, \underline{n}ot subscribing in Stage 1, and \underline{n}o action in Stage 2, is denoted as $NNN$.
Conservative consumers will only choose the behavior $NNN$. 
Among consumers who choose $PSS$ and $PDS$, a proportion $1-\alpha$ of system-incompatible consumers will abandon the software subscription in Stage 2 based on their posterior assessment.
We denote the behavior of consumers who choose the $PSS$ ($PDS$) in Stage 1 but do \underline{n}ot subscribe the software in Stage 2 as $PSN$ ($PDN$). 

Then, by following the same analytical process as in Subsection \ref{subsection:4_1} (with the detailed of analysis in \ref{appendix:US_strategy_analysis}), we have the profit of the manufacturer under $US$ strategy as
\begin{equation}
    \pi^{US}\left(r_s,p_h,p_v\right)= \left(2D_{PSS}^{US}+D_{PSN}^{US}+D_{PDS}^{US}\right)r_s+\left(D_{PSS}^{US}+D_{PSN}^{US}+D_{PDS}^{US}+D_{PDN}^{US}\right)\left(p_h-c_h\right)+2\left(p_v-c_v\right).
\end{equation}
By maximizing the profit, we get Lemma~\ref{lemma:lemma2}, with more details about closed-form solutions in Table~\ref{proof:lemma2}.

\begin{lemma}\label{lemma:lemma2}
Under $UP$ and $US$ strategies, if the cost of SSH is high ($c_h\geq v(\gamma-1)$), no consumers will adopt ADS when the initial reliability of ADS is low ($q\leq\left(c_h+2v\right)\left/\left(v+v \gamma\right)\right.$).
\end{lemma}

This lemma demonstrates the boundary conditions under which manufacturers can generate profits from ADS. 
If the development of ADS incurs high costs for manufacturers and the corresponding reliability fails to meet expectations, consumers are unlikely to derive additional utility regardless of the pricing strategy. Consequently, all consumers will choose to abandon ADS.

\subsection{$BP$ strategy}\label{subsection:4_3}
Under $BP$ strategy, the manufacturer offers the vehicle and SSH as a bundle (by price $p_b$) and sells the perpetual licensing of software (by price  $p_s$).
Progressive consumers have three potential choices for the software and the bundle of vehicle and SSH in Stage 1: (1) use ADS in two stages: \underline{p}urchasing the bundle and software in Stage 1, and \underline{h}olding them in Stage 2, is denoted as $PH$; (2) use ADS only after it upgraded: purchasing the bundle but \underline{d}elaying software purchase in Stage 1, and then \underline{p}urchasing it in Stage 2, is denoted as $DP$; (3) not to use ADS in both stages: purchasing the bundle but \underline{n}ot purchasing software in Stage 1, and \underline{n}o action in Stage 2, is denoted as $NN$.
All conservative consumers plan to choose the $NN$ behavior in Stage 1.
By Stage 2, among the progressive consumers who initially plan to \underline{d}elay software purchase, $1-\alpha$ of them find they are system-incompatible and decide \underline{n}ot to purchase, we denote the behavior of this segment of consumers as $DN$.
At the same time, some conservative consumers who have \underline{n}o willingness to purchase software will change their assessment of compatibility, and choose to \underline{p}urchase the software in Stage 2, we denote this behavior as $NP$.

By analyzing the utility associated with the different consumer choices, we can determine the number of consumers under each behavior and the resulting profit for the manufacturer as follows (the detail of analysis given in \ref{appendix:BP_strategy_analysis})
\begin{equation}
    \pi^{BP}\left(p_s,p_b\right)= \left(D_{PH}^{BP}+D_{DP}^{BP}+D_{NP}^{BP}\right)p_s+2(p_b-c),
\end{equation}
where $c=c_v+c_h$ is the total cost of vehicles and SSH. By maximizing $\pi^{BP}\left(p_s,p_b\right)$, we have Lemma~\ref{lemma:lemma3} and specific closed-form solutions in Table~\ref{proof:lemma3}.

\begin{lemma}\label{lemma:lemma3}
Under $BP$ strategy, the manufacturer's optimal price for autonomous driving software is monotonically, but not strictly, decreasing as the initial reliability of ADS ($q$) declines. When the proportion of system-compatible consumers is low ($\alpha\leq\alpha^{BP}$) and the initial reliability falls within the range $q_1^{BP}< q \leq q_2^{BP}$, the optimal price $p_s^{*}=v(\gamma-1)$ is independent of $q$.
\end{lemma}

This phenomenon arises because as $q$ decreases from a relatively high level, the utility consumers derive from ADS decreases accordingly. If the manufacturer does not adjust the price of the software in response, sales volume as well as profits will decrease. Consequently, the manufacturer will generally reduce its price as $q$ decreases.
When  $q_1^{BP} < q \leq q_2^{BP}$, the consumers who intend to use the ADS in Stage 1 will immediately purchase the software.
Because these purchases occur in Stage 1, consumers' purchasing decisions are unaffected by their posterior assessment, and consequently, the profit generated from these sales is not affected by $\alpha$.
In this scenario, any further price reduction by the manufacturer would only attract those consumers who plan to adopt the ADS in Stage 2. However, because $\alpha$ is relatively low, the incremental profit from these additional purchasers is limited. In addition, lowering the price significantly erodes the profit earned from Stage 1 purchasers. Under lower proportion of system-compatible consumers ($\alpha\leq\alpha^{BP}$), this lost profit generally outweighs the incremental gain from additional sales in Stage 2, resulting in a larger overall reduction in profit. In contrast, if the manufacturer maintains the original price, although some Stage 1 purchasers may forgo buying the software due to insufficient utility, the resulting profit loss is still smaller than the loss from a price cut. Therefore, within this range, maintaining the price is typically the superior strategy.
As $q$ continues to fall below $q_1^{BP}$, keeping the price unchanged leads to an excessive reduction in sales, substantially exacerbating the loss of profit. At this point, the manufacturer finds it more advantageous to lower the price again, aiming to offset the decline in unit profit margin through increased sales volume.

\subsection{$BS$ strategy}\label{subsection:4_4}
Under $BS$ strategy, the manufacturer sells the vehicle and SSH as a bundle (by price $p_b$) and provides software service through subscription (by price $r_s$).
Progressive consumers have three choices for the software and the bundle of vehicle and SSH: (1) use ADS in two stages: purchasing the bundle, \underline{s}ubscribing the software in Stage 1, and \underline{s}ubscribing in Stage 2, is denoted as $SS$; (2) use ADS only after it upgraded: purchasing the bundle but \underline{d}elaying the subscription in Stage 1, and \underline{s}ubscribing in Stage 2, is denoted as $DS$; (3) not to use ADS in both stages: purchasing the bundle, \underline{n}ot subscribing the software in Stage 1, and \underline{n}o action in Stage 2, is denoted as $NN$.
All conservative consumers are expected to choose $NN$ in Stage 1 based on their prior assessment of ADS compatibility.
Like consumer behavior observed in the $US$ strategy, consumers in the $BS$ strategy who initially subscribe to ADS may also choose to cancel their software subscriptions in Stage 2.
We denote the behavior of system-incompatible consumers who choose $SS$ ($DS$) in Stage 1 and \underline{n}ot to subscribe in Stage 2 as $SN$ ($DN$).
Similarly to the case in $BP$ strategy,  some conservative consumers who have \underline{n}o willingness to subscribe will change their prior assessment of compatibility, and will choose to \underline{s}ubscribe in Stage 2, and we denote this consumer behavior as $NS$.

In summary, we get the demands for the software subscription are $D_{SS}^{BS}+D_{SN}^{BS}$ in Stage 1, and $D_{SS}^{BS}+D_{DS}^{BS}+D_{NS}^{BS}$ in Stage 2 (a detailed analysis is given in Appendix~\ref{appendix:BS_strategy_analysis}).
Then, we get the manufacturer's profit as
\begin{equation}
    \pi^{BS}\left(r_s,p_b\right)= \left(2D_{SS}^{BS}+D_{SN}^{BS}+D_{DS}^{BS}+D_{NS}^{BS}\right)r_s+2(p_b-c).
\end{equation}
By maximizing $\pi^{BS}\left(r_s,p_b\right)$, we get Lemma~\ref{lemma:lemma4}, and the specific closed-form solutions are shown in Table~\ref{proof:lemma4}.
\begin{lemma}\label{lemma:lemma4}
Under the $BS$ strategy, the subscription price is set as $r_{s_h}$ when initial reliability of ADS is high ($q>q^{BS}$); and set as $r_{s_l}$ when initial reliability is low ($q\leq q^{BS}$).
Both $r_{s_h}$ and $r_{s_l}$ increase monotonically with $q$, but $r_{s_h} < r_{s_l}$.
\end{lemma}

The Lemma \ref{lemma:lemma4} presents a counterintuitive trend of the subscription price with changes in the initial reliability of ADS: the subscription price may be lower when the initial reliability is high compared to when the reliability is low.
The reason for this phenomenon is explained as follows.
When the initial reliability is low ($q \leq q^{BS}$), safety concerns discourage consumers from using ADS in Stage 1, resulting in most software subscriptions occurring in Stage 2.
In this case, the manufacturer places greater emphasis on the higher upgraded reliability in Stage 2 when setting prices $r_{s_l}$, resulting in a higher subscription price.
In contrast, when the initial reliability of ADS is high ($q > q^{BS}$),  more consumers will subscribe in Stage 1, prompting the manufacturer to decide the subscription price $r_{s_h}$ based on the initial reliability in Stage 1.
Consequently, for the same value of $q$, the subscription price $r_{s_h}$ is lower than $r_{s_l}$.

\subsection{Not all consumer purchase vehicle}\label{subsection:4_5}
In previous discussions, we analyzed the scenario in which the price of the vehicle or the bundle of the vehicle with SSH is lower than the basic utility consumers derive from the vehicle ($p_v\leq 2v $ or $ p_b\leq 2v$).
Based on this condition, all consumers will purchase the vehicle (under the unbundle strategy) or its bundle with SSH (under the bundle strategy) regardless of whether they intend to use the software.
In this section, we will discuss the cases where the condition does not hold, i.e. $p_v > 2v$ or $p_b > 2v$.
Therefore, in this scenario, the utility provided by the basic vehicle is insufficient to attract consumers relative to its price.
Only consumers who perceive ADS as highly valuable will make a purchase.
Therefore, all consumers who purchase a vehicle will inevitably choose SSH at the same time, making the unbundle strategy equivalent to the bundle strategy.

\begin{figure}[htbp]
\centering
\includegraphics[width=0.7\textwidth]{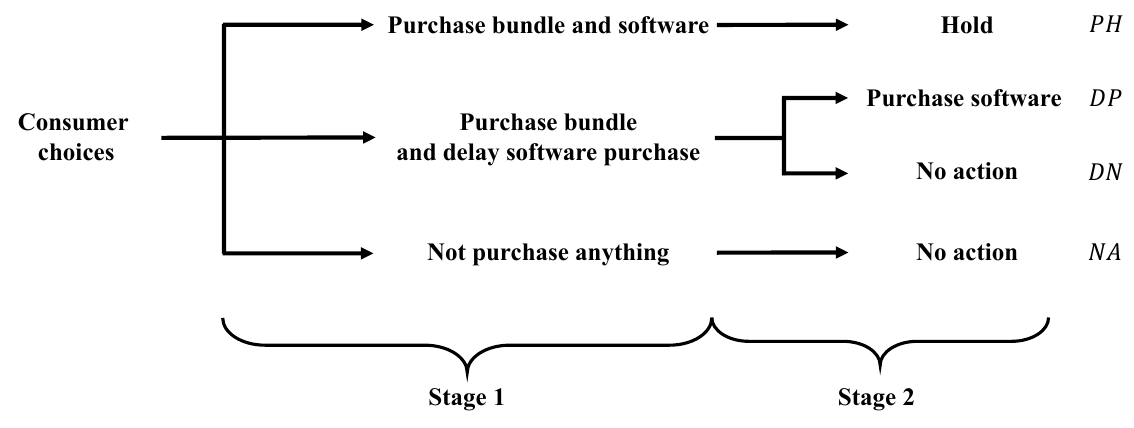}
\caption{Consumer choice under the $BP$ ($UP$) strategy when $p_b>2v$ ($p_v>2v$)}
\label{not_all_purchase}
\end{figure}

The behaviors of progressive consumers who intend to use ADS will purchase the vehicle as well as SSH, and their behavior is the same as that under $BP$ strategy: (1) \underline{p}urchasing bundle and software in Stage 1 and \underline{h}olding them in Stage 2, denoted as $PH$; (2) purchasing bundle but \underline{d}elaying the purchase of software in Stage 1, then \underline{p}urchase it in Stage 2, denoted as $DP$; (3) \underline{p}urchasing bundle in Stage 1 and \underline{n}o action in Stage 2, denoted as $PN$. 
All conservative consumers and some progressive consumers with low perceived reliability of ADS will do \underline{n}ot purchase \underline{a}nything in Stage 1 and take no action in Stage 2, which is referred to as $NA$.
Figure \ref{not_all_purchase} provides a detailed illustration of consumers' behavior selection.

Then, we give Lemma~\ref{lemma:lemma5} by comparing the manufacturer's profit in scenarios where all consumers purchase vehicles ($p_v\leq 2v$ or $p_b\leq 2v$) versus scenarios where only a subset of consumers purchase vehicles ($p_v>2v$ or $p_b>2v$).

\begin{lemma}\label{lemma:lemma5}
    Under condition $2v > c_v$, setting the price $p_v > 2v$ or $p_b > 2v$ and allowing only a portion of consumers to purchase the vehicle always results in lower profits compared to the approach that enables all consumers to purchase the vehicle.
\end{lemma}

This lemma indicates that even if ADS provides additional utility, the manufacturer is still not suggested to set a price for the base vehicle that exceeds its perceived utility.

\section{Interactions between SSH and software strategies}\label{section:discussion}
In this section, we analyze the interactions between SSH and software strategies by profit comparison. 
To simplify the parameters and enhance the visualization of the conclusions, we specify the parameters as follows, unless otherwise stated.
First, based on the report of Cinda Securities \citep{costreport}, SSH cost can be specified as $c_h = 0.1v$. 
This assignment is derived from real-world data reports and avoids the special case of no ADS purchases mentioned in the Lemma~\ref{lemma:lemma2}.
Then, from manufacturers' equilibrium profits under different strategies (shown in Tables~\ref{proof:lemma1} to~\ref{proof:lemma4}),  we find the utility of vehicle $v$ does not affect the manufacturer's decision-making.
Therefore, we set $v=1$ without loss of generality.
All proofs of Propositions~\ref{proposition:proposition1} to \ref{proposition:proposition4} given in this section are provided in Appendix~\ref{appendix:proof_proposition}.
\subsection{Impact of SSH strategy on software strategy choice}\label{subsection:subsection5_1}
By comparing the optimal profit under the $UP$ and $US$ strategies (with results shown in Tables~\ref{proof:lemma1} and ~\ref{proof:lemma2}, respectively), we get the optimal choice of the manufacturer when the vehicle and SSH are unbundled:
\begin{proposition}\label{proposition:proposition1}
Under the unbundle strategy of vehicle and SSH,
\begin{itemize}
	\item [1.] if the proportion of system-compatible consumers is low ($\alpha\leq\alpha^{U}$), the manufacturer prefers $UP$ strategy for any value of initial reliability of ADS ($q>1$);
	\item [2.] if $\alpha>\alpha^{U}$, manufacturer prefers $UP$ strategy when $q\geq q_2^{U}$ or $q\leq q_1^{U}$; otherwise ($q_1^{U} < q < q_2^{U}$), manufacturer prefers $US$ strategy.
\end{itemize}
\end{proposition}
\begin{figure}[!ht]
    \centering
	\includegraphics[width=0.6\textwidth]{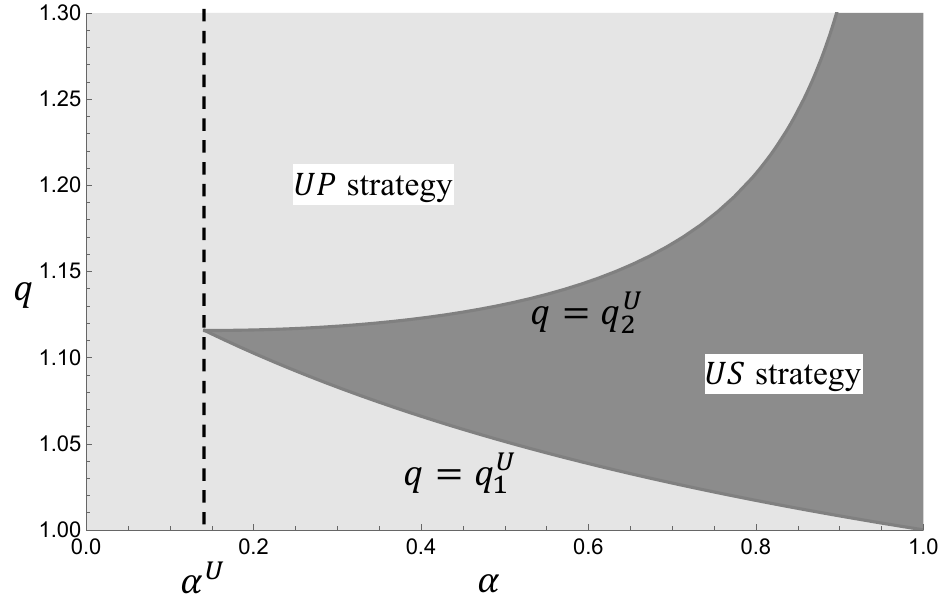}
    \caption{The manufacturer's equilibrium strategies under unbundle strategy}
    \label{proposition1}
\end{figure}

We use Figure~\ref{proposition1} to intuitively illustrate the results in Proposition~\ref{proposition:proposition1}. To better explain it, it should be first identified that, progressive consumers who are initially interested in ADS but assess it as incompatible in Stage 2 will abandon the intended software purchase or subscription. We refer to these consumers as \textit{abandon-consumers}.
Then, we highlight the distinct characteristics of the $UP$ and $US$ strategies and establish the criteria for selecting between these strategies based on their respective characteristics.

The $US$ strategy offers greater flexibility in pricing within a market with heterogeneous consumer behaviors.
Under the $US$ strategy, consumers with higher perceived reliability of ADS ($\theta$) who wish to use ADS in both stages pay a two-stage subscription, while those with lower $\theta$ who only use ADS after technological upgrade pay a one-stage fee.
Compared to the $UP$ strategy, the $US$ strategy is more effective in allowing the manufacturer to apply differentiated pricing based on consumers’ varying preferences for $\theta$, thereby capturing more consumer surplus.
We refer to this as the \textit{extra consumer-surplus effect}.
However, this characteristic of the $US$ strategy also provides a lower exit barrier for consumers, allowing them to flexibly exit the market if they find ADS is incompatible. 
In contrast, manufacturers adopting the $UP$ strategy can eliminate abandon-consumers and avoid related profit losses by packaging the sale of software with SSH (according to Lemma~\ref{lemma:lemma1}). Consumers who want to use ADS must purchase the entire package in Stage 1, with no opportunity to abandon ADS later.
We refer to this as the \textit{consumer-abandon effect}.

We now explain selection criterion between the $UP$ and $US$ strategies, which depends on the relative strength of their respective identified effects.
As the proportion of system-compatible consumers ($\alpha$) increases, the number of consumers who abandon their software subscriptions at Stage 2 under the $US$ strategy decreases.
Consequently, the strength of the consumer-abandon effect diminishes, and the extra consumer-surplus effect becomes dominant.
Therefore, the manufacturer consistently prefers the $US$ strategy as $\alpha$ increases. 
Specifically, as $\alpha$ approaches 1, no consumers will abandon their software subscriptions in Stage 2, rendering the consumer-abandon effect entirely ineffective.
In this scenario, the $US$ strategy outperforms the $UP$ strategy for any value of $q$. This validates the first conclusion in Proposition~\ref{proposition:proposition1}.

Then, to explain the second point in Proposition~\ref{proposition:proposition1}, we first introduce a new lemma.

\begin{lemma}\label{lemma:lemma6}
The proportion of consumers who choose to delay their purchase or subscription decreases with the initial reliability of the system ($q$) under the strategies $US$, $BP$, and $BS$, while no consumers delay their software purchases under the $UP$ strategy.
\end{lemma}

According to Lemma~\ref{lemma:lemma6}, the decrease in $q$ leads to a higher proportion of consumers delaying their subscription.
Since $1-\alpha$ of the delayed consumers will abandon their subscriptions in Stage 2 by finding ADS incompatible, the increase in delayed subscribers results in more abandon-consumers.
Consequently, when $q \leq q_1^{U}$, a high proportion of abandon-consumers make the consumer-abandon effect surpass the extra consumer-surplus effect, and the manufacturer prefers the $UP$ strategy.
However, this does not imply that the manufacturer always prefers the subscription strategy as \( q \) increases.
When \( q \) is extremely high (\( q \geq q_1^{U} \)), most consumers will subscribe to the software from Stage 1 under the \( US \) strategy, making consumer behaviors homogeneous and reducing the extra consumer-surplus effect.
At the same time, abandon-consumers still exist, making the consumer-abandon effect become dominant.
Only when \( q_1^{U} < q < q_2^{U} \), the proportions of consumers delaying their subscriptions and the homogeneity of consumer behavior are both relatively low, and the extra consumer-surplus effect becomes dominant and the $US$ strategy is optimal.
This supports point 2 in Proposition~\ref{proposition:proposition1}.

We now discuss the optimal choice of the manufacturer when the vehicle and SSH are sold as a bundle. By comparing the optimal profit under the $BP$ strategy and $BS$ strategy (with detailed results shown in Tables~\ref{proof:lemma3} and \ref{proof:lemma4}, respectively), we have Proposition \ref{proposition:proposition2}:
\begin{proposition}\label{proposition:proposition2}
Under the bundle strategy of vehicle and SSH,
\begin{itemize}
	\item [1.] if the proportion of system-compatible consumers is low ($\alpha\leq\alpha^{U}$), the manufacturer prefers $BP$ strategy for any value of initial reliability of ADS ($q>1$);
	\item [2.] if $\alpha>\alpha^{B}$, the manufacturer prefers to choose $BP$ strategy when $q\leq q^{B}$; otherwise ($q>q^{B}$), the manufacturer prefers $BS$ strategy.
\end{itemize}
\end{proposition}

\begin{figure}[!ht]
    \centering
	\includegraphics[width=0.6\textwidth]{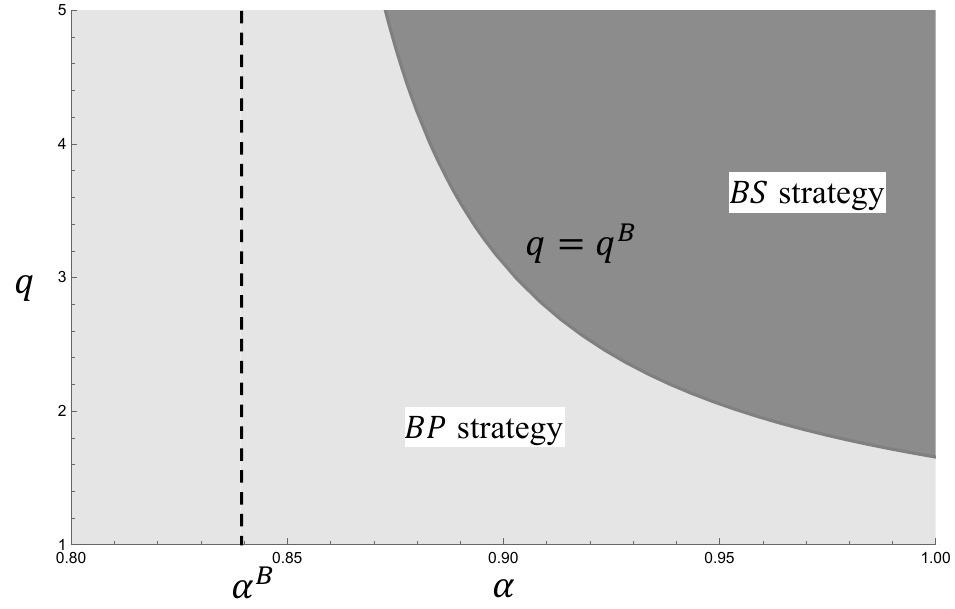}
    \caption{The manufacturer's equilibrium strategies under bundle strategy}
    \label{proposition2}
\end{figure}

To better understand this proposition, we define conservative consumers who choose to pay for ADS in Stage 2 as \textit{reentry-consumers}.
Reentry-consumers only occur under the bundle strategy as they will not purchase the SSH under the unbundle strategy and lose the opportunity to use ADS in Stage 2.
Then, we outline the characteristics of both the $BP$ and $BS$ strategies, followed by explanations of the criteria for strategy selection (shown in Figure \ref{proposition2}).

Similarly to $US$, the $BS$ strategy has an extra consumer-surplus effect as it allows manufacturers to implement more flexible pricing for different consumers.
In contrast, $BP$ strategy effectively prevents consumers who purchase the software in Stage 1 from abandoning it in Stage 2 and has the potential to achieve higher profits through consumer-abandon effect.  

When the proportion of system-compatible consumers is low ($\alpha\leq\alpha^{B}$), the number of abandon-consumers under the $BS$ strategy increases significantly, causing the $BP$ strategy with consumer-abandon effect to become dominant.
Otherwise, the initial reliability $q$ becomes a key factor influencing the manufacturer's decision. 
Generally, the manufacturer prefers the \( BS \) strategy as $q$ increases due to the following two reasons.
First, an increase in $q$ weakens the consumer-abandon effect as the high initial reliability can reduce the number of abandon-consumers.  
Second, a higher $q$ attracts more progressive consumers using ADS in Stage 1. As a result, the heterogeneity of consumer behaviors increases and the extra consumer-surplus effect is strengthened. 
The two aspects, respectively, undermine the advantages of the $BP$ strategy and strengthen the advantages of the $BS$ strategy.
From the above analysis, we conclude that when $q$ is low, the consumer-abandon effect is strong and the extra consumer-surplus effect is weak, making the $BP$ strategy profitable. Conversely, when \( q \) is high, the situation is completely opposite, which supports point 2 in Proposition~\ref{proposition:proposition2}.

From Propositions~\ref{proposition:proposition1} and~\ref{proposition:proposition2}, we observe that the perpetual licensing strategy enhances profitability through the consumer-abandon effect, while the subscription strategy enhances profitability via the extra consumer-surplus effect.
At the same time, the SSH strategy will influence the optimal software strategy choice, particularly when the initial reliability of ADS is high. 
In this case, manufacturers adopting the unbundle strategy prefer perpetual licensing, whereas those adopting the bundle strategy favor subscription.
The primary reason for this difference is that different SSH strategies alter the entry barriers for consumers in ADS market. Under the unbundle strategy, only progressive consumers enter the market, leading to highly homogeneous consumer behavior as the reliability of ADS increases, which weakens the advantage of the subscription strategy. In contrast, the bundle strategy allows conservative consumers to participate as reentry-consumers, significantly enhancing the heterogeneity of consumer behaviors and reinforcing the advantage of the subscription strategy.

\subsection{Impact of software strategy on SSH strategy choice}\label{subsection:subsection5_2}
In this section, we focus on the SSH strategy selection given software strategies. Given the perpetual licensing strategy for software, we compare the optimal profits under the $UP$ and $BP$ strategies (with results shown in Tables~\ref{proof:lemma1} and
~\ref{proof:lemma4}, respectively) and get the following proposition. 

\begin{proposition}\label{proposition:proposition3}
    Under the perpetual licensing strategy for software, the manufacturer should choose the $UP$ strategy when its initial reliability of ADS is low ($q\leq q^P$); otherwise ($q>q^P$), the $BP$ strategy is better.
\end{proposition}

\begin{figure}[!ht]
    \centering
	\includegraphics[width=0.6\textwidth]{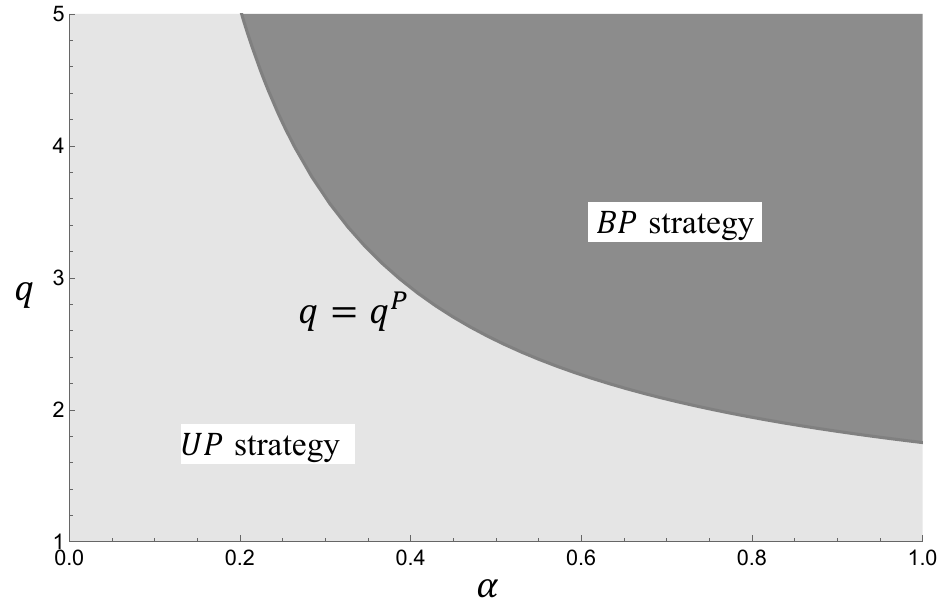}
    \caption{The manufacturer's equilibrium strategies under perpetual licensing strategy}
    \label{proposition3}
\end{figure}

Before explaining the results, we first summarize the characteristics of strategies $UP$ and $BP$.
Although consumers who intend to use ADS only after the reliability upgrade exist under both strategies, their impact on the manufacturer's profit is totally different. 
Under the $BP$ strategy, these consumers can become abandon-consumers, reducing the profit of the manufacturer.
However, under the $UP$ strategy, the manufacturer packages the software with SSH, effectively embedding all software charges into the SSH charging. Consequently, even if consumers abandon ADS usage in Stage 2, the manufacturer's profits remain unaffected.
Here, the SSH serves similarly to a deposit for the use of ADS, therefore, we refer to this phenomenon as the \textit{hardware-deposit effect}.
Although the $BP$ strategy results in profit losses from abandon-consumers, it compensates for this drawback by boosting software sales through the attraction of reentry-consumers in Stage 2, a phenomenon referred to as the \textit{consumer-reentry effect}.
The variations in the relative strength of these two effects change the optimal strategy, which is illustrated in Figure~\ref{proposition3}. We provide a detailed analysis of the effect strength that varies with different factors.

First, we find that an increase in the initial reliability of ADS ($q$) enhances the consumer-reentry effect while simultaneously weakening the hardware-deposit effect.
The enhancement of the consumer-reentry effect arises primarily because higher reliability leads to a higher software price under the $BP$ strategy (as shown in Lemma~\ref{lemma:lemma3}), allowing manufacturers to make greater profits from reentry-consumers.
Meanwhile, the proportion of consumers who delay their purchase under the $BP$ strategy gradually decreases with $q$ (according to Lemma~\ref{lemma:lemma6}), resulting in fewer abandon-consumers. Consequently, the hardware-deposit effect under $US$ strategy achieved by packaging SSH with software is weakened.

Second, an increase in the proportion of system-compatible consumers ($\alpha$) also enhances the consumer-reentry effect while simultaneously weakening the hardware-deposit effect.
Because under the $BP$ strategy, the $1-\alpha$ proportion of consumers who delay their purchase become abandon-consumers, while the $\alpha$ proportion of conservative consumers whose posterior assessment of compatibility in Stage 2 is compatible will become potential reentry-consumers. Consequently, as $\alpha$ increases, there are fewer abandon-consumers and more reentry-consumers.

From the above two points, we find that manufacturers should choose the $BP$ strategy when both $q$ and $\alpha$ are relatively high; and choose the $UP$ strategy when both $q$ and $\alpha$ are relatively low, which provides a clear explanation for the strategy regions shown in Figure~\ref{proposition3}.
Notably, when $\alpha$ approaches 1, although there are no abandon-consumers (thus completely diminishing the hardware-deposit effect) and extensive potential reentry-consumers, the $BP$ strategy does not completely dominate.
Because the higher demands generated by the $BP$ strategy come at a higher per-vehicle cost (as SSH is installed on every vehicle, regardless of whether consumers are interested in autonomous driving). 
Consequently, when the overall ADS reliability is relatively low and the revenue from selling the software does not sufficiently offset the additional SSH costs, the consumer-reentry effect can bring negative influences on the manufacturer's profit. Therefore, even in the absence of a hardware-deposit effect, the manufacturer will not adopt $BP$ strategy when the initial reliability of ADS is low.

Now, we discuss the optimal SSH strategy when manufacturers adopt a subscription strategy for software. By comparing the optimal results under the strategies $US$ and $BS$ (given in Tables~\ref{proof:lemma2} and \ref{proof:lemma4}), we have Proposition~\ref{proposition:proposition4}.

\begin{proposition}\label{proposition:proposition4}
Under the subscription strategy of software,
\begin{itemize}
	\item [1.] if the proportion of system-compatible consumers is low ($\alpha\leq\alpha^{S}$), the manufacturer prefers the $US$ strategy for any value of initial reliability of ADS ($q>1$);
	\item [2.] if $\alpha>\alpha^{S}$, the manufacturer prefers to choose the $US$ strategy when $q< q^{S}$; otherwise ($q\geq q^{S}$), the $BS$ strategy is better.
\end{itemize}
\end{proposition}

\begin{figure}[!ht]
    \centering
	\includegraphics[width=0.6\textwidth]{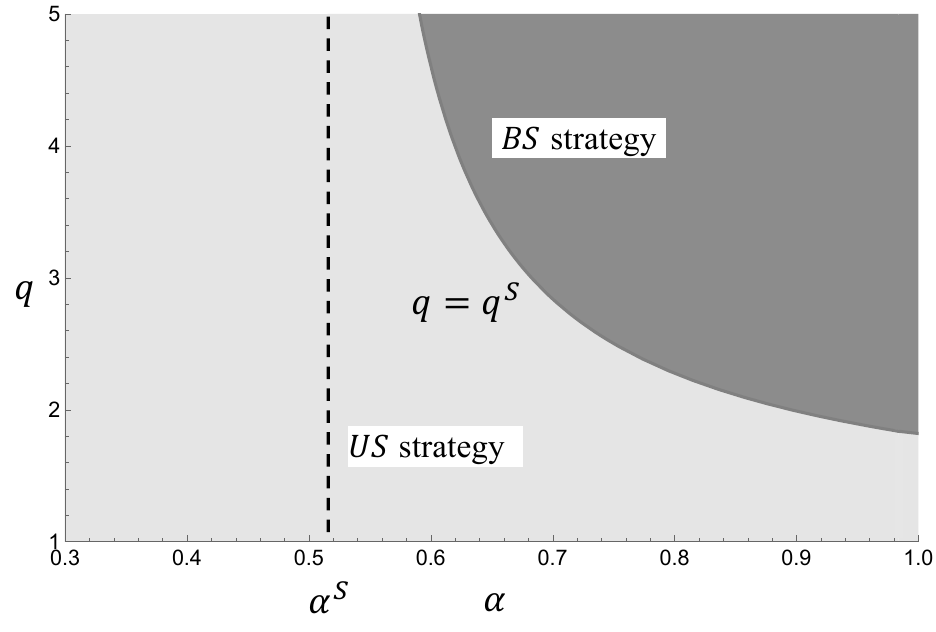}
    \caption{The manufacturer's equilibrium strategies under subscription strategy of software}
    \label{proposition4}
\end{figure}

As discussed previously, the bundle strategy facilitates reentry-consumers, which benefit the $BS$ compared to the $US$ strategy by the consumer-reentry effect. 
At the same time, the SSH sales option under the $US$ strategy helps prevent potential software revenue losses, making the hardware-deposit effect a benefit of this strategy.
Therefore, the choice between $BS$ and $US$ is determined by the relative strength of these two effects, as shown in Figure~\ref{proposition4}. Under the subscription strategy for software, these two effects are not monotonic with the initial reliability $q$, and there also exists an upper bound on $\alpha$, below which the $US$ strategy is always dominant. 

To explain the results in Figure~\ref{proposition4}, we first discuss the effect of $q$ on consumer-reentry and hardware-deposit effects in $BS$ and $US$ strategies, respectively.
The consumer-reentry effect increases in a piecewise manner with respect to $q$, with a sudden drop at the boundary between the two segments. 
Recall in Lemma~\ref{lemma:lemma4}, the subscription price $r_s$ is divided into two segments based on the initial reliability $q$ (i.e., $r_s=r_{s_h}$ when $q>q^{BS}$ and $r_s=r_{s_l}$ when $q\leq q^{BS}$).
We find both $r_s$ and the number of reentry-consumers within each segment increase with $q$, thereby strengthening the consumer-reentry effect in each segment.
However, when $q$ is relatively low, the manufacturer relies more on Stage 2 revenues and sets a higher $r_s$ based on the upgraded reliability in Stage 2. This results in a stronger consumer-reentry effect in the lower range of $q$. As $q$ increases, more progressive consumers opt to subscribe in Stage 1, prompting the manufacturer to focus more on profits in Stage 1, which weakens the consumer-reentry effect, as reentry-consumers only appear in Stage 2.
The hardware-deposit effect exhibits a trend of first weakening and then strengthening as $q$ increases.
When $q$ is relatively low, an increase in $q$ gradually reduces the proportion of consumers who delay subscribing (by Lemma~\ref{lemma:lemma6}), and diminish the hardware-deposit effect due to fewer abandon-consumers. However, once $q$ becomes sufficiently large, the number of consumers who delay their subscriptions no longer declines significantly. At this point, abandon-consumers do not vanish, as the consumers who reassess the ADS as incompatible in Stage 2 will abandon it.
These abandon-consumers cause severe profit losses as subscription prices increase with $q$ under the $BS$ strategy.
However, under the $US$ strategy, the manufacturer can shift software subscription revenues to SSH income, reducing the revenue loss.
Consequently, once $q$ is large, the hardware-deposit effect becomes significant with increases in $q$.

\begin{figure}[!ht]
    \centering
	\includegraphics[width=0.8\textwidth]{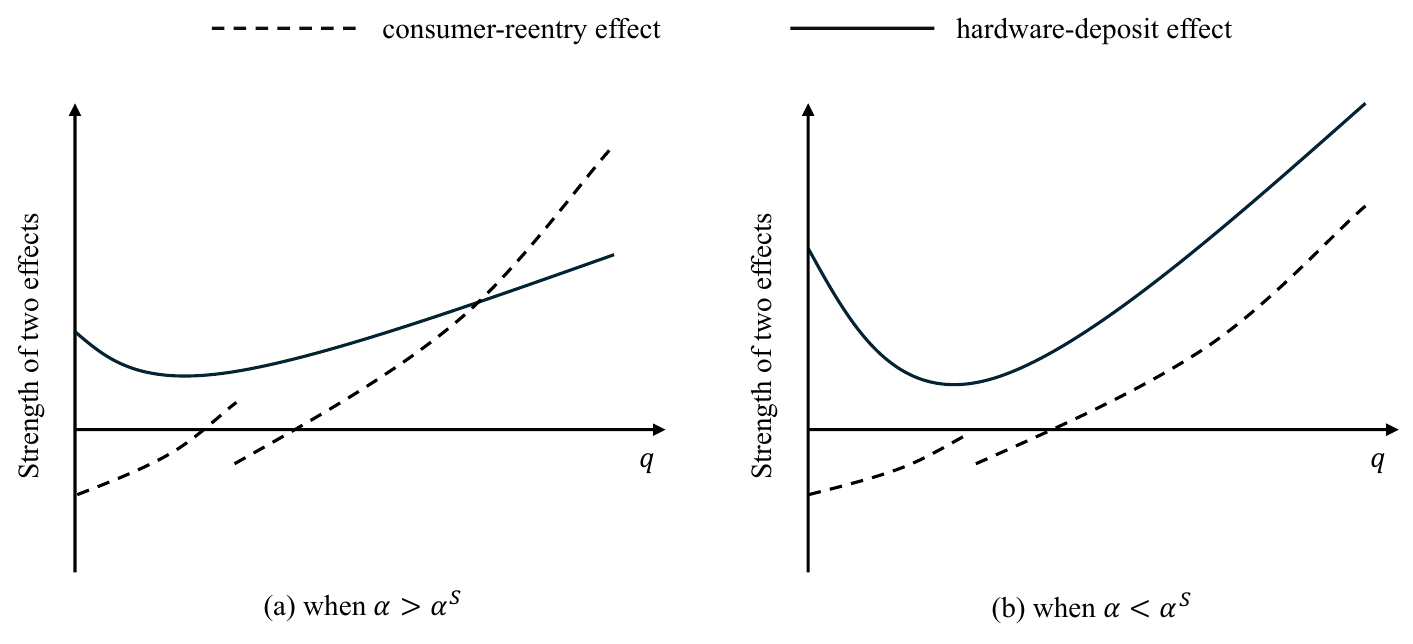}
    \caption{How the strength of the hardware-deposit effect and the consumer-reentry effect change with $q$}
    \label{proposition4_effect_change}
\end{figure}

When considering these two effects simultaneously, we find that both the consumer-reentry effect and the hardware-deposit effect intensify with further increases in $q$.
In this context, the proportion of system-compatible consumers, $\alpha$, determines which effect ultimately dominates at higher values of $q$. When $\alpha$ is relatively high, more reentry-consumers return to the market in Stage 2 and the consumer-reentry effect grows faster as $q$ increases. Meanwhile, as fewer consumers abandon their subscriptions in Stage 2, the hardware-deposit effect grows more slowly. 
In contrast, when $\alpha$ is low, exactly the opposite pattern emerges: the consumer-reentry effect expands more slowly with $q$, while the hardware-deposit effect strengthens at a faster rate.
This dynamic leads to a critical threshold, $\alpha^S$. As shown in Figure~\ref{proposition4_effect_change} (a), when $\alpha > \alpha^S$, the consumer-reentry effect ultimately becomes dominant at higher $q$, making the $BS$ strategy more appealing. In contrast, in Figure~\ref{proposition4_effect_change} (b), if $\alpha \leq \alpha^S$, the hardware-deposit effect always dominates and the manufacturer prefers the $US$ strategy.

Furthermore, when \( q \) is low, ADS provides relatively low utility, making it difficult for the additional costs of the bundle strategy to be offset by software-related profits. Hence, as illustrated in Figure~\ref{proposition4_effect_change}, the consumer-reentry effect may be negative, and its strength consistently remains below that of the hardware-deposit effect, irrespective of the value of $\alpha$.
It should be noted that the consistent dominance of the hardware-deposit effect within the low range of $q$ is a special case specific to $\gamma = 1.3$. When the upgrade level of the ADS changes, additional outcomes may emerge, which we will discuss in Section~\ref{section:extension}.

From Propositions~\ref{proposition:proposition3} and~\ref{proposition:proposition4}, we find that the selection criteria on SSH strategy are different once the software strategy changes.
The key reason for this phenomenon lies in the different exit barriers consumers face under perpetual licensing compared to subscription strategies. 
Under the perpetual licensing strategy, once consumers purchase the software, they cannot leave the market. This high exit barrier implies that as the initial reliability of ADS increases and fewer consumers delay their purchase, the manufacturer no longer needs to rely on SSH charges to secure profit, and the hardware-deposit effect of the unbundle strategy becomes weak. Consequently, the bundle strategy is more competitive when the initial reliability of ADS is high.
In contrast, under a subscription strategy, consumers can always reassess the compatibility of the ADS and exit. Due to the low exit barrier, even when the initial reliability of ADS is high and only a small fraction of consumers delay subscribing, there still exist consumers who choose to cancel their subscriptions in Stage 2, which significantly preserves the advantages of the hardware-deposit effect. As a result, the unbundle strategy becomes dominant, particularly when the proportion of consumers who are genuinely compatible with the ADS is relatively low.
\section{Extensions}\label{section:extension}
In the previous two sections, we fixed the software upgrade level at $\gamma = 1.3$, a value supported by real-world data.
In this section, we generalize this assumption to the condition $\gamma > 1 + c_h/v$, which is employed to guarantee the existence of consumers willing to purchase ADS (see Lemma~\ref{lemma:lemma2} for details), and aim to investigate the effect of $\gamma$ on the manufacturer's strategy selection.
Following the same analysis process in Section 5, we first compare the optimal software selections under different SSH strategies in Proposition~\ref{proposition:proposition5}, and then analyze the optimal SSH strategy given software strategies in Proposition~\ref{proposition:proposition6} and Observation~\ref{observation:observation1}.
The proofs of Propositions~\ref{proposition:proposition5} and \ref{proposition:proposition6} are provided in Appendix~\ref{appendix:proof_proposition}.

\begin{proposition}\label{proposition:proposition5}
Under the unbundle strategy, the mechanism for software strategy selection (between $UP$ and $US$) under different $\gamma$ is the same as Proposition~\ref{proposition:proposition1}.
Under the bundle strategy, there exists a threshold $\gamma^B$ such that the $BP$ strategy always dominates the $BS$ strategy when $\gamma \geq \gamma^B$. In other cases, the strategy selection mechanism is the same as Proposition~\ref{proposition:proposition2}.
\end{proposition}

From Proposition~\ref{proposition:proposition5}, we find that the most significant impact of $\gamma$ is that it enhances the advantage of perpetual licensing under the bundle strategy by reducing the extra consumer-surplus effect in subscription.
The reason is that a higher value of $\gamma$ causes the majority of the total utility of the use of ADS to be generated during Stage 2.
At this point, a manufacturer adopting the $BP$ strategy faces two potential options.
The first is to set a relatively low subscription price based on initial reliability to encourage early demands in Stage 1. This subscription price is substantially lower than the utility derived in Stage 2 due to the large scale of reliability upgrade. 
As a result, consumers can obtain high net utility from using ADS in Stage 2, making it difficult for manufacturers to accurately capture consumer surplus and weakening the extra consumer-surplus effect.
Alternatively, the manufacturer can set a higher subscription price based on the upgraded reliability, which encourages consumers to subscribe only in Stage 2.
In this scenario, consumer behavior becomes more homogeneous, also eliminating the extra consumer-surplus effect.
In summary, a larger $\gamma$ reduces the intensity of the extra consumer-surplus effect as well as the advantage of the subscription strategy. The manufacturer ultimately prefers perpetual licensing strategy.

It is important to note that under the bundle strategy, the perpetual licensing strategy demonstrates a more significant advantage, as the participation of reentry-consumers simultaneously increases the number of ADS users in Stage 2 and reinforces the homogeneity of consumer behavior. As a result, when $\gamma$ is sufficiently high, regardless of value of $q$ and $\alpha$, the manufacturer will choose the perpetual licensing strategy.

We now analyze the optimal SSH strategy given the software strategies. Due to the substantial differences between the two software strategies when the upgrade level $\gamma$ is varying, we first examine the strategy selection for SSH under perpetual licensing in Proposition~\ref{proposition:proposition6}.

\begin{proposition}\label{proposition:proposition6}
Under the perpetual licensing strategy for software, there exist thresholds $\gamma^P$ and $\alpha^P$, that the bundle strategy ($BP$) always dominates the unbundle strategy ($UP$) when $\gamma > \gamma^P$ and $\alpha \geq \alpha^P$. Otherwise, the strategy selection mechanism is the same as Proposition~\ref{proposition:proposition3}.
\end{proposition}

\begin{figure}[!ht]
    \centering
	\includegraphics[width=0.6\textwidth]{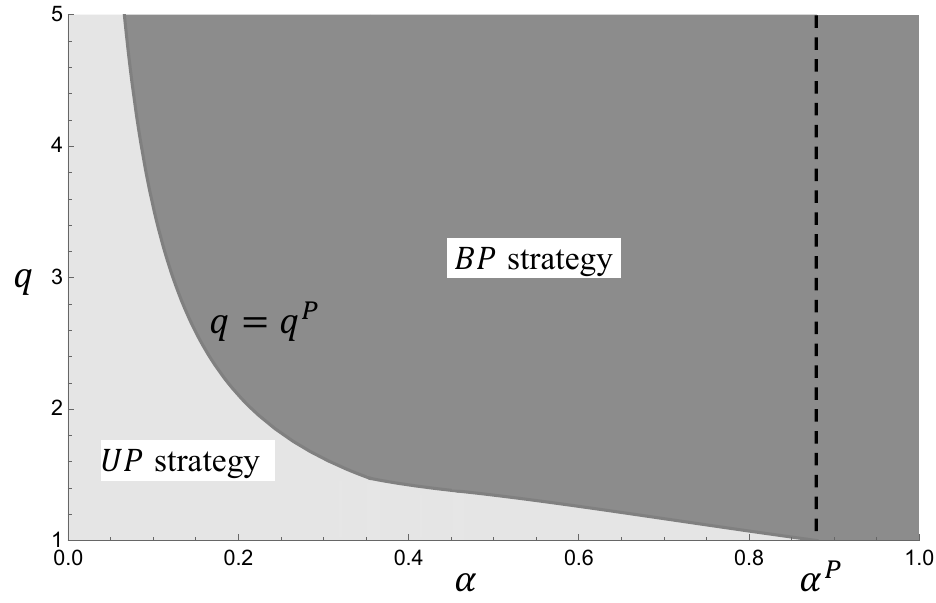}
    \caption{The manufacturer's equilibrium strategies under perpetual licensing strategy of software when $\gamma=2.5>\gamma^P$}
    \label{proposition6}
\end{figure}

Proposition~\ref{proposition:proposition6} provides the dominant condition for the $BP$ strategy based on the thresholds of $\gamma$ and $\alpha$. 
As $\gamma$ increases, the total utility consumers derive from the ADS also rises, enabling the manufacturer to set a correspondingly higher software price $p_s$. 
Meanwhile, a large $\gamma$ can narrow the relative utility gap between consumers purchasing in Stage 1 and those who delay until Stage 2, who pay the same price but receive the utility of ADS only in Stage 2. 
Consequently, more reentry-consumers are attracted to the market. Both these two factors enable the $BP$ strategy to capture additional profits from reentry-consumers, further enhancing the consumer-reentry effect.
At the same time, a larger $\alpha$ weakens the hardware-deposit effect under the $UP$ strategy due to fewer abandon-consumers.
Hence, as illustrated in Figure~\ref{proposition6}, once $\gamma$ and $\alpha$ exceed their respective thresholds (i.e., $\gamma > \gamma^P$ and $\alpha \geq \alpha^P$), the consumer-reentry effect always dominates the hardware-deposit effect, making the $BP$ strategy the optimal choice regardless of the initial ADS reliability $q$.

Finally, we analyze the optimal determination of the SSH strategy when the subscription strategy is adopted for software.

\begin{observation}\label{observation:observation1}
Under the subscription strategy, when the proportion of system-compatible consumers $\alpha$ is at a moderate level (see, between 0.6 and 0.75), as the upgrade level $\gamma$ increases, the optimal SSH strategy will alternate between the $US$ and $BS$ strategies with respect to $q$.
\end{observation}

\begin{figure}[!ht]
    \centering
	\includegraphics[width=1\textwidth]{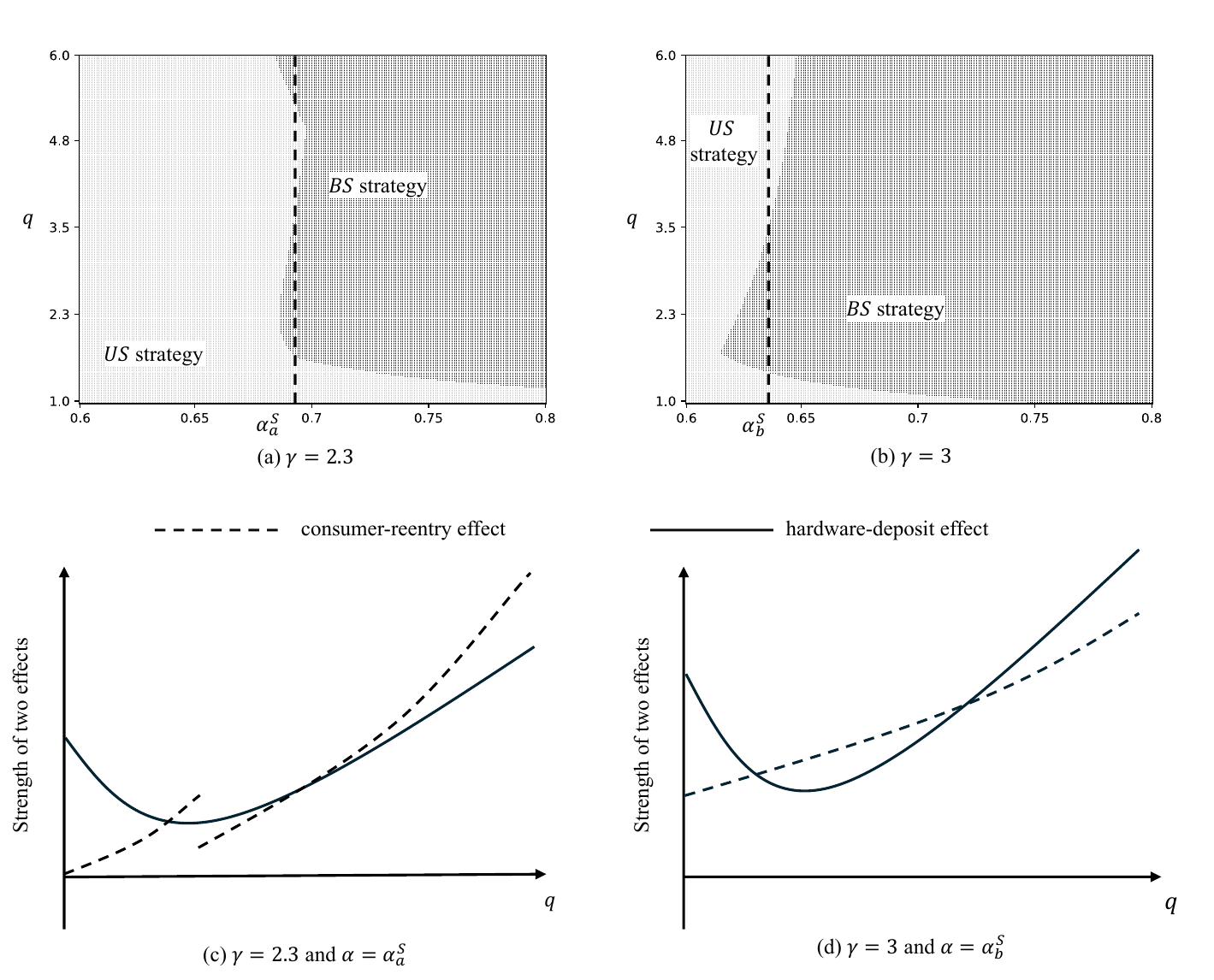}
    \caption{The manufacturer’s equilibrium strategies under the subscription strategy when $\gamma = 2.3$ and $\gamma = 3$, along with the explanation of the observed phenomenon}
    \label{figure:observation1}
\end{figure}

As mentioned in our introduction to Proposition~\ref{proposition:proposition4}, the hardware-deposit effect and consumer-reentry effect exhibit a non-monotonic trend as $q$ increases.
The consumer-reentry effect generally exhibits an increasing trend with $q$, but it undergoes a sudden decline at a certain point during this growth, and the hardware-deposit effect exhibits a trend of first weakening and then strengthening as $q$ increases.
Then, an increase in $\gamma$ enhances the utility consumers derive from using ADS, raises the software subscription price, and strengthens the consumer-reentry effect. Meanwhile, a higher upgrade level also encourages consumers to delay their subscription until Stage 2, thereby reinforcing the hardware-deposit effect.
It simultaneously enhances both the consumer-reentry effect and the hardware-deposit effect, significantly altering their relative strengths and leading to changes in the criteria for optimal strategy selection under different upgrade levels.
In Figures~\ref{figure:observation1} (a) and (b), we show the regions of optimal strategies under $\gamma = 2.3$ and $\gamma = 3$, respectively.
In Figure~\ref{figure:observation1} (a), when $\alpha = \alpha_a^S$, the $US$ and $BS$ strategies alternate as the optimal choices as $q$ increases.
To clarify the phenomenon, we present the approximate trends of the strengths of the two effects under $\gamma = 2.3$ and $\alpha = \alpha_a^S$ in Figure~\ref{figure:observation1} (c). Additionally, Figure~\ref{figure:observation1} 
(d) is provided to illustrate the phenomenon observed in Figure~\ref{figure:observation1} (b).

When $\gamma$ increases to 2.3, the enhanced overall utility of ADS enables the manufacturer to offset the additional costs of the bundle strategy through increased software subscriptions, even when the initial reliability is relatively low. Consequently, as shown in Figure~\ref{figure:observation1} (c), when $q$ increases from a low level, the reentry-consumer effect surpasses the hardware-deposit effect. However, because of a sudden decline in the reentry-consumer effect, the hardware-deposit effect temporarily regains dominance. Then, as $q$ continues to increase, the reentry-consumer effect, benefiting from its higher growth rate, eventually becomes dominant again at sufficiently high $q$. This alternating dominance of the two effects leads to the $BS$ and $US$ strategies switching back and forth as the optimal choices, which is shown in Figure~\ref{figure:observation1} (a).

When $\gamma$ increases to 3, Figure~\ref{figure:observation1} (d) shows that the reentry-consumer effect no longer experiences a sudden decline. The reason is that under the $BS$ strategy (see Table~\ref{proof:lemma4}), once the upgrade margin is sufficiently large ($\gamma > 2.5$), the manufacturer consistently sets a high subscription price and encourages all consumers to delay their subscription until Stage 2. Consequently, the reentry-consumer effect will increase monotonically with $q$ in this case. Owing to this monotonicity, when $\alpha = \alpha_b^S$, the reentry-consumer effect surpasses the hardware-deposit effect only at moderate $q$ levels, and the manufacturer opts for the bundle strategy solely at moderate values of $q$.

It is important to note that the oscillation between $US$ and $BS$ strategies primarily occurs when $\alpha$ is at an intermediate level. When $\alpha$ is relatively high, the number of abandon-consumers is small, causing the hardware-deposit effect to exhibit only a weak increasing trend with $q$, while the consumer-reentry effect consistently dominates at higher values of $q$. Similarly, when $\alpha$ is relatively low, the number of reentry-consumers is small, and the hardware-deposit effect always prevails.
However, when $\alpha$ falls in a moderate range, its influence on the relative strength of these two effects becomes less decisive, making the impact of $\gamma$ particularly crucial.
\section{Managerial Insights and Conclusions}\label{section:conclusion}
In this paper, we develop a two-stage game-theoretical model to determine the optimal business strategies and prices of each component (SSH and software) of ADS from the perspective of manufacturers. 
Considering the continuous advancement of technology and dynamic consumer behaviors, we provide the strategy selection criteria based on different development levels of ADS.
Several important results are presented as follows.

First, our study shows that SSH strategies influence the optimal software strategy by altering the entry barriers of consumers to the ADS market. We find that a newly identified extra consumer-surplus effect enhances the profitability of the subscription strategy when consumer behavior is highly heterogeneous, while a consumer-abandon effect curbs profit loss under perpetual licensing by discouraging consumer exit once a license is purchased. We further show how the choice between bundle and unbundle strategies shifts entry barriers to the ADS market, thereby altering the relative advantages of each software strategy. Under the bundle strategy, pre-installing SSH allows both progressive and conservative consumers to participate, expanding behavioral heterogeneity. When initial ADS reliability is high, progressive consumers adopt early and conservative consumers enter later, intensifying heterogeneity, magnifying the extra consumer-surplus effect, and favoring subscription. By contrast, low ADS reliability prompts both segments to delay adoption, resulting in a more homogeneous market and rendering the consumer-abandon effect dominant, making perpetual licensing more attractive. This outcome helps explain why Tesla adopts the same bundle strategy across different regions but diverges in its software strategy: deploying perpetual licensing in the Chinese market, where ADS reliability is relatively low, and offering a subscription option in North America, where reliability is higher. Under the unbundle strategy, the lack of conservative consumers limits heterogeneity, prompting manufacturers to prefer the subscription strategy only when the initial ADS reliability is at a moderate level. At this point, consumer behavior exhibits sufficient heterogeneity to capitalize on the additional consumer-surplus effect while minimizing the risk of excessive subscription cancellations.

Second, we find that software strategies influence optimal SSH strategies by changing exit barriers of consumers. Under the bundle strategy, manufacturers can raise per vehicle costs to attract more conservative consumers, thereby generating a consumer-reentry effect that encourages market expansion for greater profit. In contrast, under the unbundle strategy, manufacturers adopt a more cautious approach and rely on the hardware-deposit effect, which secures higher profits from progressive consumers without incurring additional hardware expenses. Different software strategies shift consumers’ exit barriers and thus reshape the optimal hardware decision. Under perpetual licensing, once consumers make a purchase, they cannot leave the market, resulting in high exit barriers. As ADS reliability improves, progressive consumers increasingly adopt in the early stage, and this high barrier allows the manufacturer to fully capture their profits; consequently, the hardware-deposit effect becomes less critical, and the consumer-reentry effect dominates, prompting a preference for the bundle strategy. However, when ADS reliability is low, progressive consumers delay adoption, and the potential gains from expanding into the conservative segment may be outweighed by the profit losses arising from abandoned purchases, causing the hardware-deposit effect to prevail and unbundle to be the superior option. Under the subscription strategy, consumers can easily discontinue usage if the ADS proves incompatible, resulting in low exit barriers. In most scenarios, the additional subscription volume from conservative consumers under a bundle strategy does not fully offset the cancellations among progressive consumers. Only when ADS reliability is sufficiently high and a sufficiently large fraction of users is actually compatible with the ADS, which limits cancellations among progressive consumers, does the consumer-reentry effect yield superior profits, making the bundle strategy more attractive. Under all other conditions, manufacturers favor the unbundle strategy and focus on capturing sufficient returns from progressive consumers alone.

Finally, we find that improvements in the ADS upgrade level strengthen the advantages of the perpetual licensing strategy. As upgrade levels increase, manufacturers tend to price according to the upgraded ADS performance, prompting most progressive consumers under a subscription strategy to adopt the ADS after the upgrade. This behavior leads to relatively homogeneous adoption patterns under both bundle and unbundle strategies, diminishing the extra consumer-surplus effect. Meanwhile, delayed subscriptions result in increased cancellations due to incompatibilities between consumers and the ADS, thereby strengthening the consumer-abandon effect. Together, these factors make perpetual licensing more advantageous when upgrades are substantial. In contrast, the impact of upgrade magnitude on SSH strategies is not particularly significant, as a higher upgrade magnitude simultaneously intensifies the consumer-reentry effect (price increases yield greater benefits from market expansion) and the hardware-deposit effect (delayed ADS adoption increases the likelihood of abandonment), offsetting each other’s impact.

In summary, based on the results, we provide strategy selection criteria of SSH and software for manufacturers based on initial ADS reliability and its upgrade level, which can effectively increase their profits.
\begin{itemize}
    \item [1.] For manufacturers whose ADS technology is still in its nascent stages, an unbundle SSH strategy coupled with perpetual licensing is recommended. 
    This combination maximizes profits by attracting progressive consumers, who exhibit greater openness to novel technologies and compensate for the limited technical capabilities of manufacturers.
    \item [2.] For manufacturers already offering high ADS reliability and anticipating only marginal technological advancements. It is advisable for them to conduct market research on the composition of consumers compatible with ADS before determining a strategy.
    %they the choice of strategy depends on the proportion of consumers compatible with ADS.
    \begin{itemize}
        \item [a.] If a large proportion of consumers find ADS well-compatible to their needs, combining a bundle SSH strategy with subscription is optimal, as it harnesses both progressive and conservative consumers to maximize revenue;
        \item [b.] Conversely, if a small fraction of consumers are compatible with ADS, manufacturers opting for perpetual licensing should align with a bundle approach, while those employing subscription should adopt an unbundle strategy.
    \end{itemize}
    \item [3.] For manufacturers already offering high ADS performance and anticipating significant ADS improvements in the future, bundling SSH with vehicle and offering perpetual licensing strategy is better.
\end{itemize}

Throughout our work, several potential directions remain open for future exploration. First, besides ADS, the heterogeneity in utility of basic vehicles can also be considered. Second, manufacturers can also determine the magnitude of their upgrade levels of ADS technology. It would be valuable to investigate the optimal upgrade rate  incorporates the cost of research and development (R\&D) investments in the future.
\bibliographystyle{utdref}

 \let\oldbibliography\thebibliography
 \renewcommand{\thebibliography}[1]{%
    \oldbibliography{#1}%
    \baselineskip14pt 
    \setlength{\itemsep}{10pt}
 }
\bibliography{Main}
\ECSwitch 
\ECHead{How to Promote Automated Driving with Evolving Techniques: Business Strategy and Pricing Decision} 
\section{Appendix A}\label{appendix:analysis}
This appendix provides more details on the derivation and analytics on different business strategies of ADS.
\subsection{Analysis of $UP$ strategy}\label{appendix:UP_strategy_analysis}
As given in Subsection~\ref{subsection:4_1}: $U^{PPH}=v \theta (q +\gamma q)-p_s-p_v-p_h$, $U^{PDP}=v+v \theta \gamma q -p_s-p_v-p_h$, and $U^{NNN}=2v-p_v$.
Next, let $U^{PPH} = U^{PDP}$, and by solving for $\theta$, we obtain the indifferent point between the $PPH$ and $PDP$ behaviors as $\theta_{12}^{UP}=1/q$. Similarly, we can derive the indifferent points between other behaviors, as shown in Table~\ref{indifferent_point_UB}.

\begin{table}[!ht]
\caption{Indifferent points under the $UP$ strategy}
\fontsize{10pt}{10pt}\selectfont 
   \renewcommand{\arraystretch}{1.5} 
    \begin{tabularx}{\textwidth}{>{\centering\arraybackslash}X>{\centering\arraybackslash}X}
\hline
\hline
  	Condition&Indifferent points\\
   	\hline   \\[-2.5ex]
  	$U^{PPH}=U^{PDP}$&$\theta_{12}^{UP}=\frac{1}{q}$\\[-2.5ex]\\
  	$U^{PDP}=U^{NNN}$&$\theta_{23}^{UP}=\frac{p_h+p_s+v}{\gamma  q v}$\\[-2.5ex]\\
  	$U^{PPH}=U^{NNN}$&$\theta_{13}^{UP}=\frac{p_h+p_s+2v}{\gamma  qv+qv}$\\[-2.5ex]\\
\hline
\hline
\multicolumn{2}{l}{\footnotesize
Note: indifferent points subscripts 1–3 represent $PPH$, $PDP$, and $NNN$, respectively.}
\end{tabularx}
\label{indifferent_point_UB}
\end{table}
\subsection{Analysis of $US$ strategy}\label{appendix:US_strategy_analysis}
\begin{figure}[htbp]
\centering
\includegraphics[width=0.7\textwidth]{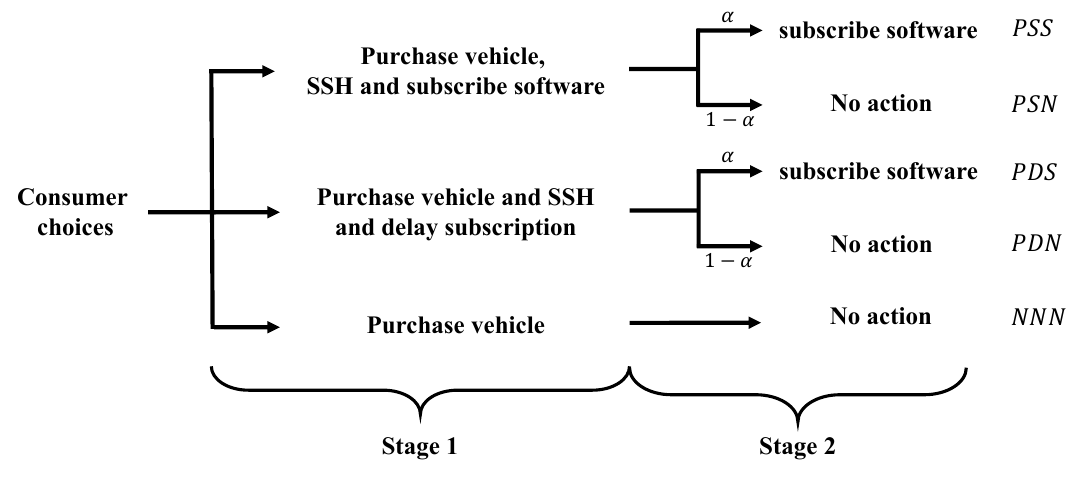}
\caption{Consumer choice under the $US$ strategy}
\label{sequence_U_S}
\end{figure}

We show all of the consumer choices in $US$ strategy in Figure \ref{sequence_U_S}.

The expected utilities of choosing $PSS$, $PDS$, and $NNN$ behaviors in Stage 1 are $U_{PSS}=(v\theta q -r_s)+(v\theta \gamma q-r_s) -p_v-p_h$, $U_{PDS}=v+v\theta \gamma q -r_s-p_v-p_h$ and $U_{NNN}=2v-p_v$, respectively.
The indifferent points for different behaviors as shown in Table \ref{indifferent_point_U_S}.
\begin{table}[!ht]
\caption{Indifferent points under the $US$ strategy}
\centering
\fontsize{10pt}{10pt}\selectfont 
   \renewcommand{\arraystretch}{1.5} 
    \begin{tabularx}{\textwidth}{>{\centering\arraybackslash}X>{\centering\arraybackslash}X}
\hline
\hline
  	Condition&Indifferent point\\
   	\hline\\[-2.5ex]
  	$U^{PSS}=U^{PDS}$&$\theta_{12}^{US}=\frac{r_s+v}{vq}$\\[-2.5ex]\\
  	$U^{PDS}=U^{NNN}$&$\theta_{23}^{US}=\frac{p_h+r_s+v}{v\gamma  q}$\\[-2.5ex]\\
  	$U^{PSS}=U^{NNN}$&$\theta_{13}^{US}=\frac{p_h+2 r_s+2v}{v\gamma  q+vq}$\\[-2.5ex]\\
\hline
\hline
\multicolumn{2}{l}{\footnotesize
Note: indifferent points subscripts 1–3 represent $PSS$, $PDS$, and $NNN$, respectively.}
\end{tabularx}
\label{indifferent_point_U_S}
\end{table}

The number of consumers finally choose $PSS$ behavior is $D_{PSS}^{US}=\alpha\cdot\text{max}\left\{0,1-\text{max}\left\{\theta_{12}^{US},\theta_{13}^{US}\right\}\right\}$, and the number of consumer finally choose the $PSN$ behavior is $D_{PSN}^{US}=(1-\alpha)\cdot\text{max}\left\{0,1-\text{max}\left\{\theta_{12}^{US},\theta_{13}^{US}\right\}\right\}$.
The number of consumers choose $PDS$ and $PDN$ behavior is $D_{PDS}^{US}=
\alpha\cdot\text{max}\left\{0,\theta_{12}^{US}-\theta_{23}^{US}\right\}$ and $D_{PDN}^{US}=
(1-\alpha)\cdot\text{max}\left\{0,\theta_{12}^{US}-\theta_{23}^{US}\right\}$, respectively.
Then, we derive the number of software subscriptions in Stage 1 equal to $D_{PSS}^{US}$, and the number of software subscriptions in Stage 2 equal to $D_{PSS}^{US}+D_{PDS}^{US}$.
The demand of SSH is $D_{PSS}^{US}+D_{PSN}^{US}+D_{PDS}^{UBS}+D_{PDN}^{US}$, and all consumers will purchase the vehicle.
Hence, the profit of the manufacture is
\begin{equation}
    \pi^{US}\left(r_s,p_h,p_v\right)= \left(2D_{PSS}^{US}+D_{PSN}^{US}+D_{PDS}^{US}\right)r_s+\left(D_{PSS}^{UBS}+D_{PSN}^{US}+D_{PDS}^{US}+D_{PDN}^{US}\right)\left(p_h-c_h\right)+2\left(p_v-c_v\right).
\end{equation}
\subsection{Analysis of $BP$ strategy}\label{appendix:BP_strategy_analysis}
\begin{figure}[htbp]
\centering
\includegraphics[width=0.7\textwidth]{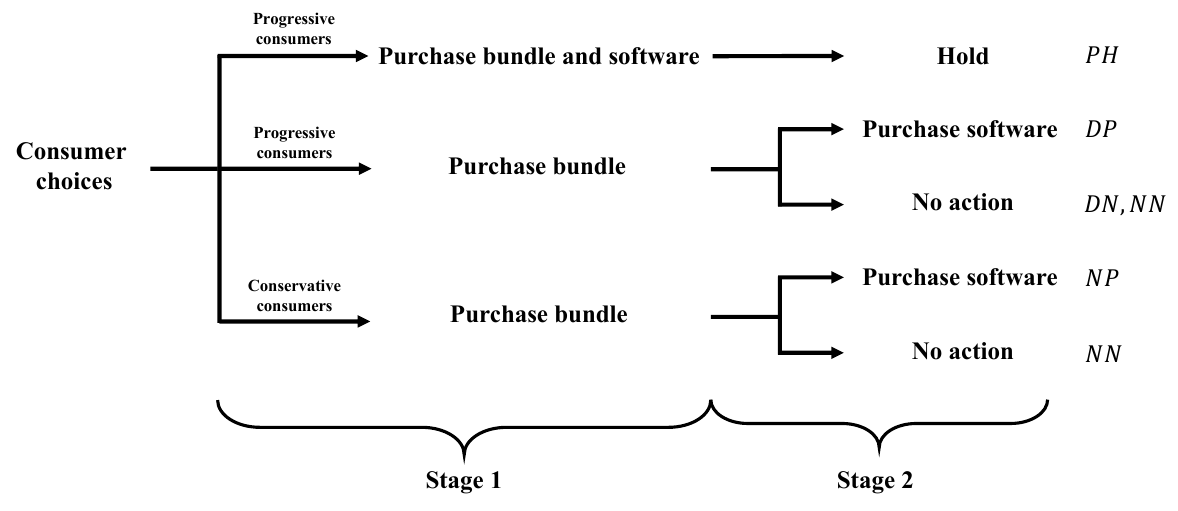}
\caption{Consumer choice under the $BP$ strategy}
\label{sequence_B_P}
\end{figure}

The consumer choice under $BP$ strategy is presented in Figure \ref{sequence_B_P}.

If consumers choose $PH$ behavior, their expected utility functions in Stage 1 are $U^{PH}= v\theta (q +\gamma q)-p_s-p_b$.
The expected utility functions of other choices are $U^{DP}=v+ v\theta \gamma q-p_s-p_b$, $U^{NN}=v+v -p_b$.
For conservative consumers, the utility of using the software in Stage 2 is $U^{NP}=v+ v\theta \gamma q-p_s-p_b$.
conservative consumers should make a decision between $NP$ and $NN$ in Stage 2.
Then, we get the indifferent points for different choices as shown in Table \ref{indifferent_point_B}.

\begin{table}[!ht]
\caption{Indifferent points under the $BP$ strategy}
\centering
\fontsize{10pt}{10pt}\selectfont 
   \renewcommand{\arraystretch}{1.5} 
    \begin{tabularx}{\textwidth}{>{\centering\arraybackslash}X>{\centering\arraybackslash}X}
\hline
\hline
  	Condition&Indifferent point\\
   	\hline\\[-2.5ex]
  	$U^{PH}=U^{DP}$&$\theta_{12}^{BP}=\frac{1}{q}$\\[-2.5ex]\\
  	$U^{DP}=U^{NN}$&$\theta_{23}^{BP}=\frac{p_s+v}{\gamma  qv}$\\[-2.5ex]\\
  	$U^{PH}=U^{NN}$&$\theta_{13}^{BP}=\frac{p_s+2v}{\gamma  qv+qv}$\\[-2.5ex]\\
        $U^{NP}=U^{NN}$&$\theta_{34}^{BP}=\frac{p_s+v}{\gamma  qv}$\\[-2.5ex]\\
\hline
\hline
\multicolumn{2}{l}{\footnotesize
Note: indifferent point subscripts 1–4 represent $PH$, $DP$, $NN$, and $NP$, respectively}
\end{tabularx}
\label{indifferent_point_B}
\end{table}

The number of consumers choosing different behaviors is determined through these indifferent points.
For progressive consumers, the number of consumer finally choose $PH$ behavior is $D_{PH}^{BP}=\text{max}\left\{0,1-\text{max}\left\{\theta_{12}^{BP},\theta_{13}^{BP}\right\}\right\}$, the number of consumer choice $DP$ behavior is $D_{DP}^{BP}=\alpha\cdot\text{max}\left\{0,\theta_{12}^{BP}-\theta_{23}^{BP}\right\}$, and the number of $DN$ behavior is $D_{DN}^{BP}=(1-\alpha)\cdot\text{max}\left\{0,\theta_{12}^{BP}-\theta_{23}^{BP}\right\}$.
For conservative consumers, the number of consumer choice $NP$ behavior is $D_{NP}^{BP}=\alpha\cdot\text{max}\left\{0,1-\theta_{34}^{BP}\right\}$.
Then, the sales volume of software in two stages is $D_{PH}^{BP}+D_{DP}^{BP}+D_{NP}^{BP}$, and the sales volume of the bundle of the vehicle and SSH is 2. 
Therefore, we have the total profit of the manufacturer as follows.
\begin{equation}
    \pi^{BP}\left(p_s,p_b\right)= \left(D_{PH}^{BP}+D_{DP}^{BP}+D_{NP}^{BP}\right)p_s+2(p_b-c),
\end{equation}
where $c=c_v+c_h$ is the total cost of vehicles and SSH.
\subsection{Analysis of $BS$ strategy}\label{appendix:BS_strategy_analysis}
\begin{figure}[ht!]
\centering
\includegraphics[width=0.7\textwidth]{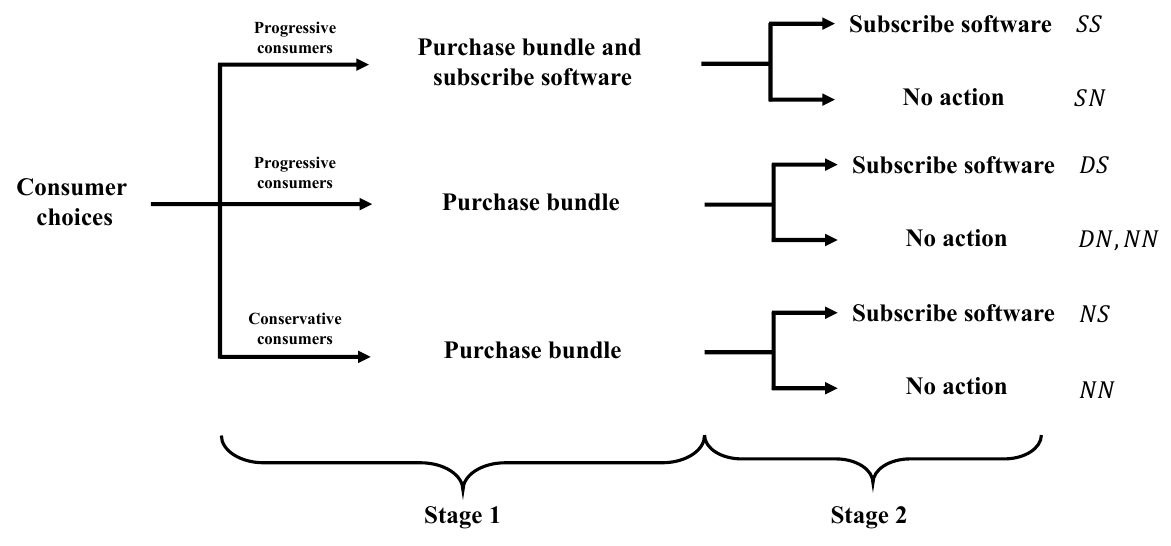}
\caption{Consumer choice in the $BS$ strategy}
\label{sequence_B_S}
\end{figure}

The consumer choice is shown in Figure \ref{sequence_B_S}.
The expected utility functions for different behaviors in Stage 1 are $U^{SS}= v\theta (q +\gamma q)-p_s-p_b$, $U^{DS}=U^{NS}=v+ v\theta \gamma q-p_s-p_b$, $U^{NN}=v+v -p_b$.
The indifferent points for different behaviors as shown in Table~\ref{indifferent_point_BS}.
\begin{table}[!ht]
\caption{Indifferent points under the $BS$ strategy}
\centering
\fontsize{10pt}{10pt}\selectfont 
   \renewcommand{\arraystretch}{1.5} 
    \begin{tabularx}{\textwidth}{>{\centering\arraybackslash}X>{\centering\arraybackslash}X}
\hline
\hline
  	Condition&Indifferent point\\
   	\hline\\[-2.5ex]
  	$U^{SS}=U^{DS}$&$\theta_{12}^{BS}=\frac{r_s+v}{qv}$\\[-2.5ex]\\
  	$U^{DS}=U^{NN}$&$\theta_{23}^{BS}=\frac{r_s+v}{\gamma  q\gamma v}$\\[-2.5ex]\\
  	$U^{SS}=U^{NN}$&$\theta_{13}^{BS}=\frac{2r_s+2v}{qv+\gamma  qv}$\\[-2.5ex]\\
        $U^{NS}=U^{NN}$&$\theta_{34}^{BS}=\frac{r_s+v}{\gamma  q\gamma v}$\\[-2.5ex]\\
   [0.5ex]
\hline
\hline
\multicolumn{2}{l}{\footnotesize
Note: indifferent point subscripts 1–4 represent $SS$, $DS$, $NN$, and $NS$, respectively}
\end{tabularx}
\label{indifferent_point_BS}
\end{table}

The number of consumers choosing different behaviors is determined through these indifferent points.
For progressive consumers, the number of consumer finally choose $SS$ behavior is $D_{SS}^{BS}=\alpha\cdot \text{max}\left\{0,1-\text{max}\left\{\theta_{12}^{BS},\theta_{13}^{BS}\right\}\right\}$.
The number of consumer finally choose $SN$ behavior is $D_{SN}^{BS}=(1-\alpha)\cdot\text{max}\left\{0,1-\text{max}\left\{\theta_{12}^{BS},\theta_{13}^{BS}\right\}\right\}$.
The number of consumer choice $DS$ behavior is $D_{DS}^{BS}=\alpha\cdot\text{max}\left\{0,\theta_{12}^{BS}-\theta_{23}^{BS}\right\}$, and the number of $DN$ behavior is $D_{DN}^{BS}=(1-\alpha)\cdot\text{max}\left\{0,\theta_{12}^{BS}-\theta_{23}^{BS}\right\}$.
For conservative consumers, the number of consumer choice $NS$ behavior is $D_{NS}^{BS}=\alpha\cdot\text{max}\left\{0,1-\theta_{34}^{BS}\right\}$.
Then, the number of software subscriptions in Stage 1 is $D_{SS}^{BS}+D_{SN}^{BS}$, and the number of software subscriptions in Stage 2 is $D_{SS}^{BS}+D_{DS}^{BS}+D_{SN}^{BS}$.
All consumers purchase the bundle of the vehicle and SSH, and the sales volume is 2. 
Therefore, we have the total profit of the manufacturer as following
\begin{equation}
    \pi^{BS}\left(r_s,p_b\right)= \left(2D_{SS}^{BS}+D_{SN}^{BS}+D_{DS}^{BS}+D_{NS}^{BS}\right)p_s+2(p_b-c),
\end{equation}
where $c=c_v+c_h$ is the total cost of vehicles and SSH.
\section{Appendix B}\label{appendix:proof_lemma}
This appendix includes proofs of all Lemmas.
\subsection{Proof of Lemma~\ref{lemma:lemma1}}\label{appendix_proof:lemma1}
\begin{table}[!ht] 
    \caption{The manufacturer's optimal pricing and profit under $UP$ strategy}
    \fontsize{10pt}{10pt}\selectfont 
    \centering 
    {\def\arraystretch{1.5}
  \begin{tabular*}{\textwidth}{@{\extracolsep{\fill}}cccc}
\hline
\hline
   Case&Conditions &Price&Profit\\
\hline\\[-2ex]
   1&$\gamma\geq\gamma^{UP} , q\leq q_1^{UP} $&$p_h^{*}=\frac{1}{2} (t_h+\gamma  q-1)v,p_s^{*}=0,p_v^{*}=2v$&$\frac{(t_h-\gamma  q+1)^2v}{4 \gamma  q}+2(2v-c_v)$\\[-2ex]\\
   2&$\left\{\begin{array}{cc}
   	\gamma\geq\gamma^{UP},& q>q_1^{UP}\\
   	\gamma\leq\gamma^{UP},& q>q_2^{UP}\\
   \end{array}\right.$&$p_h^{*}+p_s^{*}=\frac{1}{2}(t_h+\gamma  q+q-2)v, p_v^{*}=2v$&$\frac{(\gamma  q+q-2-t_h)^2v}{4 (\gamma +1) q}+2(2v-c_v)$\\[-2ex]\\
    3&$\gamma\leq\gamma^{UP}, q\leq q_2^{UP} $&$p_h^{*}+p_s^{*}=(\gamma  q+q-2)v,p_v^{*}=2v$&$2(2v-c_v)$\\[-2ex]\\
\hline
\hline
\\[-6ex]\\
    \multicolumn{4}{l}{\footnotesize Note:
$q_1^{UP}=\frac{\gamma ^2+\gamma +\sqrt{\gamma  (\gamma +1) (\gamma -t_h-1)^2}}{\gamma ^2+\gamma }$, $q_2^{UP}=\frac{t_h+2}{\gamma +1}$ and  $\gamma^{UP} =1+t_h$.}
    \end{tabular*}
    }

    \label{proof:lemma1}
\end{table}
\subsection{Proof of Lemma~\ref{lemma:lemma2}}\label{appendix_proof:lemmach}
\begin{table}[!ht] 
    \fontsize{10pt}{10pt}\selectfont 
    \centering 
        \caption{The manufacturer's optimal pricing and profit under $US$ strategy}
        {\def\arraystretch{1.5}  %change the number for increasing or decreasing the spacing.
            %\scriptsize %uncomment this for changing the font size
\begin{tabular*}{1\textwidth}{@{\extracolsep{\fill}}cccc} 
\hline
\hline
   Case&Conditions &Price&Profit\\
\hline\\[-2ex]

   1& $\left\{\begin{array}{ccc}
   	\gamma\geq \gamma^{US}, &\alpha\leq\alpha^{US}, &q\leq q_1^{US} \\
   	\gamma\geq \gamma^{US},&\alpha>\alpha^{US}, & q\leq q_2^{US}
   \end{array}\right.$&$\begin{matrix}
   	p_h^{*}=\frac{1}{2} (-\alpha -1) \epsilon_1+\frac{1}{2} v (\gamma  q+t_h-1),\\
   	r_s^{*}=\epsilon_1,p_v^{*}=2v\\
   \end{matrix}$&$\frac{(t_h-\gamma  q+1)^2v}{4 \gamma  q}+2(2v-c_v)-\text{O}(\epsilon_1)$\\[-2ex]\\
    2&$\gamma\geq \gamma^{US} ,\alpha>\alpha^{US} ,q_2^{US}<q\leq q_3^{US}$&$p_h^{*}=p_h^{US},r_s^{*}=r_s^{US},p_v^{*}=2v$&$\pi_2^{US}$\\[-2ex]\\
   3&$\left\{\begin{array}{ccc}
   	\gamma\geq \gamma^{US} ,&\alpha\leq \alpha^{US} ,&q\geq q_1^{US}\\
   	\gamma\geq \gamma^{US} ,&\alpha>\alpha^{US} ,&q\geq q_3^{US}\\
   	\gamma\leq\gamma^{US} ,& &q>q_4^{US}\\
   \end{array}\right.$&$\begin{matrix}
   	p_h^{*}=\frac{\epsilon_1}{2} (-\alpha -3)+\frac{1}{2} (t_h+\gamma  q+q-2)v,\\
   	r_s^{*}=\epsilon_1,p_v^{*}=2v\\
   \end{matrix}$&$\frac{(-t_h+\gamma  q+q-2)^2v}{4 (\gamma +1) q}+2\left(2v-c_v\right)-\text{O}(\epsilon_1)$\\[-2ex]\\
   4&$\gamma\leq\gamma^{US}, q\leq q_4^{US} $&$p_h^{*}=(\gamma -1) q v,r_s^{*}=(q-1) v,p_v^{*}=2v$&$2(2v-c_v)$\\[-2ex]\\

\hline
\hline
\\[-6ex]\\
    \multicolumn{4}{l}{\footnotesize Note:
$\gamma^{US}=1+t_h$, $\alpha^{US}=2 \gamma -2 \sqrt{\gamma  (\gamma +1)}+1$, $q_1^{US}=\frac{\gamma ^2+\sqrt{\gamma  (\gamma +1) (-t_h+\gamma -1)^2}+\gamma }{\gamma ^2+\gamma }$, $q_2^{US}=\frac{\alpha +(\alpha -1) t_h+2 \gamma -1}{(\alpha +1) \gamma }$,}\\[-1.8ex]\\
    \multicolumn{4}{l}{\footnotesize
$q_3^{US}=\frac{\left(\alpha ^2-1\right) t_h (\gamma +1)-2 \left(-\left(\alpha ^2 (\gamma +1)\right)-\alpha  \gamma ^2+\alpha +\gamma ^2+\sqrt{(\alpha -1) (\gamma +1) \left(\alpha ^2-2 \alpha -4 \gamma +1\right) (t_h-\gamma +1)^2}+\gamma \right)}{(\alpha -1) (\gamma +1) (\alpha  \gamma +\alpha +3 \gamma -1)}$,$q_4^{US}=\frac{t_h+2}{\gamma +1}$,}\\[-1.8ex]\\
    \multicolumn{4}{l}{\footnotesize
$p_h^{US}=\frac{v \left(-\alpha  \gamma +\alpha +\alpha ^2 (\gamma  q-1)+\gamma  (-2 \gamma  q+q+1)-t_h (\alpha +2 \gamma -1)\right)}{\alpha ^2-2 \alpha -4 \gamma +1}$, 
$r_s^{US}=\frac{v (\alpha +2 \gamma -\alpha  \gamma  q-\gamma  q+(\alpha -1) t_h-1)}{\alpha ^2-2 \alpha -4 \gamma +1}$,}\\[-1.8ex]\\
    \multicolumn{4}{l}{\footnotesize
$\pi_2^{US}=-\frac{v \left(\alpha +\gamma +\gamma  q^2 (\alpha +\gamma )-q (\alpha  \gamma +\alpha +3 \gamma +t_h (\alpha +2 \gamma -1)-1)+t_h^2+\alpha  t_h+t_h\right)}{q \left(\alpha ^2-2 \alpha -4 \gamma +1\right)}+2(2v-c_v)$.}
    \end{tabular*}
    }
    \label{proof:lemma2}
\end{table}
\subsection{Proof of Lemma~\ref{lemma:lemma3}}\label{appendix_proof:lemma3}
\begin{table}[!ht] 
        \fontsize{10pt}{10pt}\selectfont 
    \centering 
            \caption{manufacturer's optimal pricing and profit under $BP$ strategy}
        {\def\arraystretch{1.5}  %change the number for increasing or decreasing the spacing.
            %\scriptsize %uncomment this for changing the font size
\begin{tabular*}{1\textwidth}{@{\extracolsep{\fill}}cccc} 
\hline
\hline
   Case&Conditions &Price&Profit\\
\hline\\[-2ex]
    1&$\left\{\begin{matrix}
        \alpha\leq\alpha^{BP}, q\leq q_1^{BP}\\
   	\alpha>\alpha^{BP} , q\leq q_3^{BP}\\
   \end{matrix}\right.$&$p_{s_1}=\frac{v(\alpha  (\gamma +\gamma  q-2)+\gamma  (q-1))}{4 \alpha },p_b^{*}=2v$&$\frac{(\alpha  (\gamma +\gamma  q-2)+\gamma  (q-1))^2v}{8 \alpha  \gamma  q}+2(2v-c)$\\[-2ex]\\
   2&$\alpha\leq \alpha^{BP}, q_1^{BP}<q\leq q_2^{BP}$&$p_{s_2}=v(\gamma-1),p_b^{*}=2v$&$\frac{(\alpha +1) (\gamma -1) (q-1)v}{q}+2(2v-c)$\\[-2ex]\\
   3& $\left\{\begin{matrix}
   	\alpha>\alpha^{BP} , q>q_3^{BP}\\
   	\alpha\leq \alpha^{BP}, q>q_2^{BP}
   \end{matrix}\right.$&$p_{s_3}=\frac{v(\alpha  (\gamma +1) (\gamma  q-1)+\gamma  (\gamma  q+q-2))}{2 (\alpha  \gamma +\alpha +\gamma )},p_b^{*}=2v$&$\frac{(\alpha  (\gamma +1) (\gamma  q-1)+\gamma  (\gamma  q+q-2))^2v}{4 \gamma  (\gamma +1) q (\alpha  \gamma +\alpha +\gamma )}+2(2v-c)$\\[-2ex]\\
\hline
\hline
\\[-6ex]\\
    \multicolumn{4}{l}{\footnotesize
Note: $\alpha^{BP}=\frac{\gamma }{\gamma +1}$, $q_1^{BP}=\frac{\alpha  (3 \gamma -2)+\gamma }{(\alpha +1) \gamma },q_2^{BP}=\frac{\alpha  \left(2 \gamma ^2+\gamma -1\right)+2 \gamma ^2}{(\alpha +1) \gamma  (\gamma +1)}$,} \\[-1.8ex]\\\multicolumn{4}{l}{\footnotesize$q_3^{BP}=\frac{\alpha ^3 (\gamma +1)^2 \gamma ^2+\alpha ^2 \left(\gamma ^2+3 \gamma +2\right) \gamma ^2+\sqrt{2} \sqrt{\alpha  (\alpha +1)^2 (\gamma -1)^2 \gamma ^2 (\gamma +1) (\alpha  \gamma +\alpha -\gamma )^2 (\alpha  \gamma +\alpha +\gamma )}+\alpha  \left(\gamma ^2-\gamma ^4\right)-\gamma ^4-\gamma ^3}{(\alpha +1)^2 \gamma ^2 (\gamma +1) (\alpha  \gamma +\alpha -\gamma )}$}
    \end{tabular*}
    }

    \label{proof:lemma3}
\end{table}
\subsection{Proof of Lemma~\ref{lemma:lemma4}}\label{appendix_proof:lemma4}
\begin{table}[ht] 
    \centering 
            \fontsize{10pt}{10pt}\selectfont 
                \caption{The manufacturer's optimal pricing and profit under $BS$ strategy}
        {\def\arraystretch{1.5}  %change the number for increasing or decreasing the spacing.
            %\scriptsize %uncomment this for changing the font size
\begin{tabular*}{1\textwidth}{@{\extracolsep{\fill}}cccc} 
\hline
\hline
   Case&Conditions &Price&Profit\\
\hline\\[-2ex]
   1&$\left\{\begin{array}{ccc}
   	\gamma\leq\frac{5}{2},&0<\alpha<1,&q\leq q^{BS} \\
   	\gamma>\frac{5}{2},&0<\alpha<\alpha^{BS}, &q\leq q^{BS} \\
    \gamma>\frac{5}{2},&\alpha^{BS}\leq\alpha<1,&q>1
   \end{array}\right.$
   &$r_{s_l}=\frac{1}{2} v(\gamma  q-1),p_b^{*}=2v$&$\frac{\alpha  (\gamma  q-1)^2v}{2 \gamma  q}+2(2v-c)$\\[-2ex]\\
   2& $\left\{\begin{array}{ccc}
   	\gamma\leq\frac{5}{2},&0<\alpha<1,&q> q^{BS} \\
   	\gamma>\frac{5}{2},&0<\alpha<\alpha^{BS},& q> q^{BS}
   \end{array}\right.$&$r_{s_h}=\frac{v (2 \alpha  (\gamma  q-1)+\gamma  (q-1))}{2 (2 \alpha +\gamma )},p_b^{*}=2v$&$\frac{(2 \alpha  (\gamma  q-1)+\gamma  (q-1))^2v}{4 \gamma  q (2 \alpha +\gamma )}+2(2v-c)$\\[-2ex]\\
\hline
\hline
\\[-6ex]\\
    \multicolumn{4}{l}{\footnotesize
Note: $\alpha^{BS}=\frac{1}{2 (\gamma -2)}$, $q^{BS}=\frac{\sqrt{2} \sqrt{\alpha  (\gamma -1)^2 (2 \alpha +\gamma )}+2 \alpha +\gamma }{\gamma -2 \alpha  (\gamma -2) \gamma }$.}
    \end{tabular*}
    }

    \label{proof:lemma4}
\end{table}
\subsection{Proof of Lemma~\ref{lemma:lemma5}}\label{appendix_proof:lemma5}
The proof process is omitted.
\subsection{Proof of Lemma~\ref{lemma:lemma6}}\label{appendix_proof:lemma6}
The proof process is omitted.
\section{Appendix C}\label{appendix:proof_proposition}
This appendix includes proofs of all Propositions.
In the subsequent proofs, we will substitute the baseline utility of vehicle usage as $v = 1$ to simplify the complexity of the calculations. As introduced in Section~\ref{section:discussion}, this substitution will not affect the conclusions.
\subsection{Proof of Proposition~\ref{proposition:proposition1}}\label{appendix_proof:proposition1}
The proof process is omitted.
\subsection{Proof of Proposition~\ref{proposition:proposition2}}\label{appendix_proof:proposition2}
The proof process is omitted.
\subsection{Proof of Proposition~\ref{proposition:proposition3}}\label{appendix_proof:proposition3}
The proof process is omitted.
\subsection{Proof of Proposition~\ref{proposition:proposition4}}\label{appendix_proof:proposition4}
The proof process is omitted.
\subsection{Proof of Proposition~\ref{proposition:proposition5}}\label{appendix_proof:proposition5}
The proof process is omitted.
\subsection{Proof of Proposition~\ref{proposition:proposition6}}\label{appendix_proof:proposition6}
The proof process is omitted.
\end{document}